\documentclass[twocolumn]{autart}
\usepackage[latin1]{inputenc}
\usepackage[english]{babel}
\usepackage{amsmath,amssymb}
\usepackage{graphicx,graphicx,verbatim,subfigure}
\usepackage{hyperref}
\usepackage{amssymb,latexsym,amsmath} 
\usepackage{color} 

\newtheorem{theorem}{Theorem}

\newtheorem{assumption}{Assumption}

\newtheorem{defi}{Definition}

\newtheorem{exx}{Example}
\newtheorem{remm}{Remark}

\newenvironment{proofof}{\noindent {\em Proof of }}{\hfill \hspace*{1pt}
\hfill $\square$}

\newenvironment{remark}{\begin{remm}\rm }{\hfill \hspace*{1pt} \hfill $\lrcorner$\end{remm}}

\newcommand{\realn}{\real^n}

\newcommand\diag{{\rm diag}} 

\newcommand\sgn{{\rm sgn}}

\newcommand\real{\ensuremath{{\mathbb R}}}

\newcommand{\setof}[1]{\ensuremath{\left \{ #1 \right \}}}

\newcommand{\tto}{\ensuremath{\rightrightarrows}}

\newcommand\mymatrix[2]{\left[\begin{array}{#1} #2 \end{array}\right]}
\newcommand{\smallmat}[1]{\left[ \begin{smallmatrix}#1
    \end{smallmatrix} \right]}

\newcommand\dz{\mathrm{dz}}

\newcommand{\dom}{\mathrm{dom}}

\begin{document}

\begin{frontmatter}
\title{{ Event-triggered transmission for linear control \\ over communication channels }}

\author[liegea]{Fulvio Forni\thanksref{liege}}
\author[romaa]{Sergio Galeani\thanksref{roma}}
\author[melba]{Dragan Ne\v{s}i\'{c}\thanksref{melb}}
\author[laas,trento]{Luca Zaccarian\thanksref{laas_trento}}

\address[liegea]{Department of Electrical Engineering and Computer Science,
Universit\'e de Li{\`e}ge, 4000 Li{\`e}ge, Belgium.\\[-.6cm]}
\address[romaa]{DICII, University of Roma, Tor Vergata, 
Via del Politecnico 1, 
00133 Roma, Italy\\[-.6cm]}
\address[melba]{EEE Department, University of Melbourne, Australia.\\[-.6cm]}
\address[laas]{CNRS, LAAS, 7 avenue du colonel Roche, F-31400 Toulouse, France and
Univ. de Toulouse, LAAS, F-31400 Toulouse, France\\[-.6cm]}        
\address[trento]{Dipartimento di Ingegneria Industriale, University of
  Trento, Italy}        
\thanks[liege]{This paper presents research results of the Belgian Network DYSCO
(Dynamical Systems, Control, and Optimization), funded by the
Interuniversity Attraction Poles Programme, initiated by the Belgian
State, Science Policy Office. The scientific responsibility rests with
its authors. Work supported by FNRS. {\tt fforni@ulg.ac.be}}                  
\thanks[roma]{Work supported by ENEA-Euratom and MIUR. {\tt galeani@disp.uniroma2.it}}
\thanks[melb]{Work supported by the Australian Research Council under the Future Fellowship. {\tt d.nesic@ee.unimelb.edu.au}} 
\thanks[laas_trento]{Work supported by HYCON2 Network of Excellence ``Highly-Complex
and Networked Control Systems'', grant agreement 257462 and by the ANR project LimICoS, contract number 12 BS03 005 01.} 

\begin{abstract}
We consider an exponentially stable closed loop interconnection of a continuous linear plant 
and a continuous linear controller, and we study the problem of interconnecting the plant 
output to the controller input through a { digital channel.}
We propose a family of ``transmission-lazy'' sensors 
whose goal is to transmit the measured plant output information as little as possible 
while preserving closed-loop stability. In particular, we propose two transmission 
policies, providing conditions on the transmission parameters. 
These guarantee global asymptotic stability when the plant state is available or
when an estimate of the state is available (provided by a classical 
continuous linear observer). Moreover, under a specific condition, 
they guarantee global exponential stability. \vspace{-5mm}
\end{abstract}

\end{frontmatter}

\date{\today}

\maketitle

\section{Introduction}
\label{sec:introduction}

In recent years, much attention has been devoted to the study of
{ closed-loop control systems interconnected by a digital
  channel
where the information transmission is triggered by specific
event-triggered conditions.}
The interest in this class of control systems is motivated by 
the increased computational capability
required by control and estimation algorithms in addition to the presence
of emerging control applications 
{ wherein
the actuators of a control system may be non-colocated with the
sensing devices (e.g, drilling systems, remote handling systems) so
that the plant output is collected by the controller via a digital
channel which may have some stringent bandwidth requirement. 
This class of systems is a very specific subclass of the much more
general topic of networked control systems (see, e.g., the recent surveys
\cite{YangIEE06,HespanhaIEEE07} and references therein).
 Indeed, while
in general networked control systems, various subcomponents 
are spread over a wide territory or are technologically built in such
a way that several subcomponents of the control system communicate
over shared and low capacity digital channels,
the study of event-triggered and self-triggered systems
\cite{Anta10,Carnevale07,Cervin08,Mazo08,LpNCS,Tabuada07,Wang08a,Wang08b,Zhang01}
led to a significant amount of research results where the core problem
under consideration is that of two nodes (the sensing node and the
actuating one) communicating through
a (low capacity) digital channel where the transmission policy is
determined based on suitable Lyapunov-like conditions involving some
(more or less coarse) measurement of the plant state.}

{ A natural way to represent and suitably write the dynamics of 
this specific two-nodes configuration}
is to use the hybrid systems notation, namely a
state-space description wherein the state flows according to some
continuous-time rules and, at some specific times, called jump times,
it jumps following some discrete-time jump rule. A framework for the
representation of hybrid systems that has been recently proposed in
\cite{Goebel06,Goebel04} allows for a quite natural description of
these phenomena with useful Lyapunov like results that have been
proven to apply to large classes of systems described using this
framework (see, e.g., \cite{CaiTAC07,CaiTAC08} and the survey
\cite{GoebelCSM09}).
This framework was used in connection with networked control systems
in \cite{Carnevale07,LpNCS}, and recently in \cite{Postoyan11a,Postoyan11b},
where Lyapunov tools are used to 
model ISS properties of networked control systems and the 
MATI (maximum allowable transfer interval), to preserve asymptotic stability.

\begin{figure}[ht!]
\centering
\includegraphics[width=0.96\columnwidth]{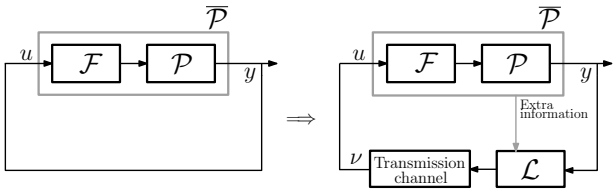}
 	\caption{Nominal closed-loop $\mathcal{S}$ and 
transmission-lazy closed-loop system $\mathcal{S}_{\mathcal{L}}$.}
 	\label{fig:SL}
\end{figure}

Here, we consider a closed-loop system that consists of
a linear controller driving a linear plant to guarantee
closed-loop asymptotic stability, as shown in Figure~\ref{fig:SL},
and we break the continuity of the transmission of the measured
plant output $y$ to the controller input $u$ by 
introducing the \emph{transmission-lazy sensors}, devices which
measure the output $y$ and decide whether or not sending
this measurement to the controller input $u$ through a transmission channel, based on non-periodic
Lyapunov-based policies. We call these sensors ``transmission-lazy'' to 
resemble the fact that their goal is to avoid transmitting too often,
so as to keep the digital channel load small enough.
We suppose that each sensor is able to perform some computation
on the measured plant output and, possibly,
on extra available signals.
{
The ideal scenario where this approach is relevant corresponds to
cases where due to some technological constraint, there is a
transmission line between a location where all the sensors are
installed and a second location where the actuators are placed
with a transmission channel inbetween (see Figure~\ref{fig:SL}).}

The contribution of this paper consists in casting 
the above problem within the hybrid framework summarized in \cite{GoebelCSM09}
and proposing two transmission policies for the transmission-lazy sensors
which preserve the (global exponential) stability of the original closed-loop
system. { This result is achieved without requiring any modification
to the design of the original controller}. 
For simplicity, we first consider two transmission 
policies based on the state of the plant and the
measurement error through a suitable Lyapunov-like function:\\
\textbullet~a \emph{synchronous transmission policy} 
 where each sensor is aware of the conditions
 of the other sensors so that a transmission 
 is a global decision { of the sensing node}. In this case, the sensors 
 transmit a new sample all together when some suitable condition occurs;\\
\textbullet~an \emph{asynchronous transmission policy} 
 where each sensor knows its own measurement error and the state of
 the plant, { which is available in the sensing node}. Then,
 it decides autonomously { (namely, without any information on the measurement 
 error of the  other sensors) } whether or not to transmit a new sample.\\
Then, we remove the dependence from the state by showing that the closed-loop
results achieved by the transmission-lazy sensors are preserved when the information
on the state (state-feedback) is replaced by an estimate 
from an observer (output feedback) { located in the sensing
  node.
To this aim, the adopted hybrid formulation is a fundamental tool.} 

Within the existing literature, the results in this
paper can be seen as a specific application of the hybrid
framework \cite{GoebelCSM09} to a peculiar control problem.
In this sense, our paper
is a constructive solution
along the general lines of \cite{Carnevale07,LpNCS,Postoyan11a},
where Lyapunov tools and the hybrid framework of
\cite{GoebelCSM09} are used { in similar 
contexts.
Moreover, the motivation behind our work is that
of event-triggered sampling where
many interesting results have been published in recent years (see
in \cite{Cervin08,Mazo08,Tabuada07,Wang08a} and 
references therein)}. {
Additional work sharing the scenario of Figure~\ref{fig:SL}
is that of \cite{SharonTAC12,NesicTAC09} and references therein, where 
the feedback signal is affected by an undesired quantization effect,
rather than the presence of the communication channel.
Within the event-triggered sampling context,}
taking into account linear 
systems, our work complements \cite{Tabuada07,Wang08a}, by
casting similar problems and approaches within the hybrid systems framework, and
proposing asynchronous transmission policies which are 
different from the synchronous ones considered in
\cite{Tabuada07,Wang08a}. 
{
We show in the paper how restricting the attention to
linear systems (whereas \cite{Tabuada07,Wang08a} considers nonlinear
systems) allows us to design transmission policies which lead to
improved results, both in terms of architectures and of 
achievable performance, as compared to those in  \cite{Tabuada07,Wang08a}
where, since a much more general nonlinear scenario is considered, the
results obtained are more conservative.
Finally,} this work extends the results
proposed in \cite{Forni10} by enforcing a dwell-time between
transmissions and by introducing exponential bounds on the
asymptotic stability guaranteed by the transmission policies.
Finally, practical stability results of the output feedback case
in \cite{Forni10} are now replaced by asymptotic (exponential) 
stability results.

The paper is structured as follows. 
In Section \ref{sec:preliminaries} we 
introduce the notation and give some preliminaries 
on hybrid systems. In Section \ref{sec:model}
we introduce the problem data. Then, in Sections 
\ref{sec:synchronous}, \ref{sec:asynchronous}, and
\ref{sec:output_feedback} we illustrate the two policies
first using information from the state of the plant, 
then relaxing this requirement by introducing an observer.
Simulation examples are given in Section \ref{sec:example}.

\section{Notation and preliminaries}
\label{sec:preliminaries}
Given a vector $v$, $v^T$ denotes the transpose vector 
of $v$. Given two vectors $w$ and $v$, $\langle v,w \rangle = w^Tv$. 
Given a set $a=\{a_1,\dots,a_n\}$ where $a_i\in \real$ for each
$i=1\,\dots,n$, $\mathrm{diag}(a)$ denotes 
a diagonal matrix having the entries of $a$ 
on the main diagonal. Both the Euclidean norm of a vector 
and the corresponding induced matrix norm
are denoted by $|\cdot|$. 
For a vector $v\in \real^n$ and a set $\mathcal{A}\subset \realn$ 
$|v|_{\mathcal{A}}:=\inf_{y\in \mathcal{A}}|y-v|$.
{ 
Given a set $\mathcal{A}\in\realn$, the set $\mathcal{A}+\varepsilon\mathbb{B}$
$\varepsilon\geq 0$, 
is the set of vectors
$v$ such that $|v|_{\mathcal{A}} \leq \varepsilon$. }
A {continuous} function
$\alpha:\real_{\geq 0}\rightarrow\real_{\geq 0}$ is said 
to belong to class $\mathcal{K}$ if it is strictly increasing and $\alpha(0) = 0$;
it belongs to class $\mathcal{K}_{\infty}$ if, moreover, $\lim_{r \rightarrow+\infty}\alpha(r)=+\infty$.
For any $s\in\real$, consider the function 
$f\,:\, \real \to \real$ defined by 
$f(s) = 0$ if $|s|\leq 1$, and 
$f(s) = \sgn(s)(|s|-1)$ if $|s|\geq 1$. Then,
for any $s=\smallmat{s_1 & \dots & s_n}^T\in\realn$,
the deadzone function $\dz\,:\,\realn \to \realn$ is 
given by $\dz(s)=\diag( f(s_1),\dots, f(s_n))$.

We summarize next the essential notation associated with the hybrid 
systems framework, outlined in \cite{Goebel04}, 
for which several results have been developed in
\cite{Goebel06,SanfeliceTAC07,Sanfelice08} and partially 
summarized in \cite{GoebelCSM09}. 
A hybrid system $\mathcal{H}$ 
is a tuple $(\mathcal{C},\mathcal{D},F,G)$, where $\mathcal{C}\subseteq\realn$ and $\mathcal{D}\subseteq\realn$ 
are, respectively, the \emph{flow set} and the \emph{jump set}, 
while $F:\realn\tto\realn$ and $G:\realn\tto\realn$ are set-valued
mappings, called \emph{flow map} and 
\emph{jump map}, respectively.
$F$ and $G$ characterize the continuous and the discrete 
evolution of the system, while $\mathcal{C}$ and $\mathcal{D}$ 
characterize subsets of $\realn$ 
where such evolution may occur. A hybrid system is usually
represented as follows
\begin{equation}
 \mathcal{H} =
 \left\{
 \begin{array}{rl} 
  \dot{x} \in F(x) & \quad x\in \mathcal{C} \\
  x^+ \!\! \in G(x) & \quad x\in \mathcal{D}. \\
 \end{array}
 \right. \label{sys}
\end{equation}
Intuitively, the state continuously flows through $\mathcal{C}$, 
by following the dynamics given by $F$, or it jumps from $\mathcal{D}$, 
according to $G$. This hybrid evolution of
the system can be conveniently characterized by using the
notion of \emph{hybrid time domain} which is a subset
$E$ of $\real_{\geq0} \times \mathbb{N}$ given by
the union of infinitely many intervals of the form
$[t_j,t_{j+1}]\times\{j\}$ where $0=t_0\leq t_1 \leq t_2 \leq,\dots$, 
or of finitely many such intervals, with the last one possibly of the form
$[t_j,t_{j+1}]\times\{j\}$, $[t_j,t_{j+1})\times\{j\}$, or
$[t_j,\infty]\times\{j\}$. Considering the notion of 
\emph{hybrid arc} $x:\mathrm{dom}\, x \to \realn$ 
given by
(i)~$\mathrm{dom}\, x$ is a hybrid time domain and 
(ii)~for each $j$, the function $t\mapsto x(t,j)$ is a locally absolutely 
continuous function on the interval $I_j=\setof{t\,:\,(t,j)\in\mathrm{dom}\,x}$,
we can define a \emph{solution} to a hybrid system $\mathcal{H}$ 
as a hybrid arc which satisfies the following two conditions
(i)~for each $j\in\mathbb{N}$ such that $I_j$ has a nonempty interior
\begin{equation*}
 \label{eq:hybsol_CF}
\begin{array}{ll} 
 \dot{x}(t,j) \in F(x(t,j)) & \mbox{ for almost all } t \in I_j \\ 
 x(t,j) \in \mathcal{C} & \mbox{ for all } t\in [\min I_j, \sup I_j),
\end{array}
\end{equation*}
and (ii)~for each $(t,j) \in \mathrm{dom}\,x$ such that $(t,j+1)\in \mathrm{dom}\, x$,
\begin{equation*}
\label{eq:hybsol_DG} 
\begin{array}{rcl} 
 x(t,j+1) & \in & G(x(t,j)) \\
 x(t,j)   &\in & \mathcal{D}. 
\end{array} 
\end{equation*} 
Solutions to hybrid systems may exist for a finite time,
due to the constraints on the state motion enforced by the $\mathcal{C}$ and $\mathcal{D}$
sets. We say that a solution $x$ is \emph{maximal} if there does not exists $x'$
such that $x$ is a truncation of $x'$ to some proper subset of
$\dom\,x'$. We say that a solution $x$ is 
\emph{complete} if $\mathrm{dom}\,x$ is unbounded.

{

Hybrid system $\mathcal{H}$ satisfies the 
\emph{basic conditions} \cite{GoebelCSM09},\cite{Goebel12} if \\
\textbullet~$\mathcal{C}$ and $\mathcal{D}$ are closed sets in $\realn$; \\
\textbullet~$F:\realn \tto \realn$ is an 
outer semicontinuous\footnote{We recall here that a 
set valued mapping is outer semicontinuous if its graph is a closed set.
Note that for single valued functions $f:\realn \to \realn$, outer semicontinuity is
equivalent to continuity.} set-valued mapping, locally bounded on $\mathcal{C}$, and 
$F(x)$ is nonempty and convex for each $x\in\mathcal{C}$; \\
\textbullet~$G:\realn \tto \realn$ is an outer semicontinuous set-valued mapping, 
locally bounded on $\mathcal{D}$, and such that $G(x)$ is nonempty for each $x \in\mathcal{D}$.\\
These conditions are fundamental to guarantee robustness of the stability
results presented in this paper.
}

Finally, following \cite{GoebelCSM09}, for a hybrid system $\mathcal{H}$
the set $\mathcal{A}$ is 
(i)~\emph{stable} if for each $\epsilon > 0$ 
there exists $\delta > 0$ such that any solution $x$ to $\mathcal{H}$ 
with $|x(0,0)|_\mathcal{A}\leq \delta$ satisfies $|x(t, j)|_\mathcal{A}\leq\epsilon$ 
for all $(t, j)\in \mathrm{dom}\, x$; 
(ii)~\emph{attractive} if { every maximal solution is complete} and 
there exists $\delta > 0$
such that any solution $x$ to $\mathcal{H}$ with $|x(0,0)|_\mathcal{A}\leq \delta$
is bounded and $|x(t,j)|_\mathcal{A} \to 0$ as $t+j\to \infty$, 
whenever $x$ is complete; 
(iii)~\emph{asymptotically stable} if it is both stable and attractive;
(iv)~\emph{exponentially stable} if for some $\gamma>0$ and $k>0$, 
each solution $x$ to $\mathcal{H}$ satisfies 
$|x(t,j)|_{\mathcal{A}} \leq k \exp(-\gamma(t+j)) |x(0,0)|_{\mathcal{A}}$ for all $(t,j) \!\in\! \dom\, x$. 
For an asymptotically (exponentially) stable compact set $\mathcal{A}$, 
the \emph{basin of attraction} $\mathcal{B}_{\mathcal{A}}$ is the set of points
in $\realn$ from which each solution is bounded and the complete
solutions converge to ${\mathcal{A}}$. Finally, 
if $\mathcal{B}_{\mathcal{A}}=\realn$ then ${\mathcal{A}}$ 
is \emph{globally asymptotically (exponentially) stable}.

\section{The transmission-lazy closed-loop system} 
\label{sec:model}
Consider a \emph{nominal closed-loop system}, $\mathcal{S}$, defined by  
the cascade interconnection $\overline{\mathcal{P}}$ 
of a \emph{linear controller} $\mathcal{F}$ and a \emph{linear plant} 
$\mathcal{P}$, given by
\begin{equation}
 \overline{\mathcal{P}}\,:\,\left\{
	\begin{array}{rcl}
	 \dot{x} & = & Ax+Bu \\
			y	& = & Cx,
	\end{array}
\right.\label{eq:plant}
\end{equation}
where $A,B,C$ are matrices, $x\in\realn$, and $u,y\in\real^q$, 
and by the interconnection relation $u=y$ 
between the controller input and the plant measured output, 
from which we get
\begin{equation}
 \mathcal{S}\,:\,\left\{
	\begin{array}{rcl}
	 \dot{x} & = & \left(A+BC\right)x \\
			y	& = & Cx.
	\end{array}
\right.\label{eq:nominal_cls}
\end{equation}
We consider the following standing assumption.
\begin{assumption} 
\label{assume:well-posed}
 The nominal closed-loop system $\mathcal{S}$ 
is exponentially stable.
\end{assumption}
Consider now the introduction of a new device $\mathcal{L}$, 
the \emph{transmission-lazy sensors}, or {\em t-lazy sensors} within the feedback interconnection, 
as shown in Figure \ref{fig:SL}. These intelligent sensors
monitor the measured output $y$ and decide autonomously
when transmitting a new sample of the output, denoted by $\nu$,  
to the control input $u$ 
with the twofold goal of preserving the stability of the
closed-loop system while breaking the continuity on the transmission
of the measured plant output.
Looking at Figure \ref{fig:SL}, the arising \emph{transmission-lazy closed-loop system} $\mathcal{S}_\mathcal{L}$,
namely the closed-loop system of \eqref{eq:plant} through the 
interconnection $u = \nu$, is a hybrid system which
combines together the
continuous dynamics of the plant-controller cascade 
$\overline{\mathcal{P}}$ and the discrete dynamics of 
the t-lazy sensors $\mathcal{L}$. Its continuous dynamics can be modeled by 
\begin{equation}
\label{eq:lazy_sys_flow}
	\left\{ \begin{array}{lll}
		\dot{x} &=& Ax + B \nu  \\
		\dot{\nu} &=& 0, \\
	 \end{array}\right.  
\end{equation}
where $x$ takes into account the plant-controller 
cascade dynamics while $\nu\in\real^q$ denotes the \emph{state} of the
t-lazy sensors, (each element $\nu_i$ of $\nu$ is 
related to the $i$th-sensor), which replaces 
the interconnection $u=y$ of $\mathcal{S}$ with $u=\nu$, 
which is kept constant during flows by $\dot{\nu} = 0$. The discrete dynamics is given by
\begin{equation}
\label{eq:lazy_sys_jump}
	\left\{ \begin{array}{lll}
		x^+ &=& x  \\
		\nu^+ &=& g(x,\nu,\tau), \\
	\end{array}\right.  
\end{equation}
where $x$ does not change during jumps 
while $\nu$ is updated to $g(x,\nu,\tau)$, 
$g:\realn\times\real^q\times\real^p\to \real^q$,
whose definition is given next and represent a transmission 
($\tau\in \real^p$ is an external timer characterized below). 

Finally, we equip the t-lazy sensors with a bounded timer to guarantee 
a non-zero dwell-time between updates, whose dynamics is given by
\begin{equation}
\label{eq:lazy_sys_timer}
	\left\{ \begin{array}{llll}
		\dot{\tau}_i &=& 1-\dz(\frac{\tau_i}{\rho} ) 
		& \quad\tau \geq 0 \\
		\tau_i^+ &=& 0 
                & \quad\tau \geq \Delta,\\
	\end{array}\right. 
\end{equation}
where $0<\Delta< \rho$ are design parameters, 
which guarantee 
that $\tau_i$ is bounded by $2\rho$, it has
rate $1$ for $\tau_i \leq \rho$, and it
may be reset to zero only if $\tau_i \geq \Delta$.

In what follows we will consider two scenarios in which 
either (i)~one timer $\tau\in\real$ is shared among sensors, 
i.e. $p=1$ (synchronous policy), or 
(ii)~each $i$th sensor has its own timer $\tau_i$.
i.e. $p=q$ (asynchronous policy). Thus, given 
$\smallmat{\tau_1 & \dots & \tau_p}^T\in\real^p$, 
$\overline{1} := \smallmat{1 & \dots & 1}^T\in\real^p$,
and a function $h:\realn\times\real^q\times\real^p\to\real^p$
which represents possible asynchronous resets of timers, 
the hybrid dynamics of the transmission-lazy closed-loop system
(or {\em t-lazy closed loop}) can be summarized as follows: 
\begin{equation}
\label{eq:lazy_sys}
 \mathcal{S}_\mathcal{L}:
  \begin{array}{ll} 
	\left\{ \begin{array}{lll}
		\dot{x} &=& Ax + B \nu  \\
		\dot{\nu} &=& 0 \\
		\dot{\tau} &=& \overline{1}-\dz(\frac{\tau}{\rho} ) \\
	 \end{array}\right. & 
	\quad(x,\nu,\tau)\in \mathcal{C}_\Delta \\
	\left\{\begin{array}{lll}
		x^+ &=& x  \\
		\nu^+ &=& g(x,\nu,\tau) \\
		\tau^+ &=& h(x,\nu,\tau) \\
	\end{array} \right. & 
        \quad(x,\nu,\tau)\in \mathcal{D}_\Delta \\
	 \begin{array}{lll}
		y &=& Cx.  \\
	 \end{array} 
 \end{array}
\end{equation}
Within the model proposed in \eqref{eq:lazy_sys},
the sets $\mathcal{C}_\Delta$ and $\mathcal{D}_\Delta$ 
and the parameter $\Delta$ and $\rho$
will be designed to decide whether or not to update $\nu$. Therefore,
a \emph{transmission policy} is given by the tuple
$(\mathcal{C}_\Delta,\mathcal{D}_\Delta,\Delta,\rho)$.

{
\begin{remark}
 The transmission of a sample is modeled in \eqref{eq:lazy_sys}
 as an instantaneous reset of the value $\nu$,
 which will typically assign to $\nu$ the current value of
 the output $y$. The proposed sample transmission model is a
 rough abstraction of a (possibly convoluted) process where the
 overall dynamics of the 
 { transmission channel} plays a fundamental role. For example,
 the model does not consider transmission delays, noise corruption of
 the samples, packet drop, and many other features
 of { digital transmission}. While 
 these phenomena concur to the evaluation
 of the closed-loop performance, within a certain
 bounded magnitude, they will not affect the main stability results 
 established below. In particular, relying on the robustness to
 small perturbations guaranteed by the hybrid framework \cite{Goebel04}, 
 the stability of our closed loop still holds
 in the presence of (small) transmission noise and delays
 (see Section \ref{sec:robustness}).
 \end{remark}
}

\section{State feedback: synchronous transmission}
\label{sec:synchronous}

\subsection{The error dynamics}
We consider a \emph{synchronous} transmission policy
in which the transmission of the samples is a global decision 
based on the knowledge of $x$ and $\nu$ and of a timer state 
shared among sensors, { i.e. $p=1$. 
The whole t-lazy sensors state $\nu$ is updated at once,
that is $\nu=g(x,\nu,\tau) := Cx = y$. Moreover, 
at updates, the timer state is reset to zero, that is,
$h(x,\nu,\tau)=0$.}

For simplicity of the exposition
we consider the coordinate transformation $(x,e) :=(x,\nu-y)$,
from which we can rewrite the t-lazy closed loop \eqref{eq:lazy_sys} 
as follows
\begin{equation}
\label{eq:elazy_sys}
\begin{array}{ll}
	 \left\{ \begin{array}{lll}
	 \dot{x} & = & F_{11}x + F_{12}e  \\
  	 \dot{e} & = & F_{21}x  +F_{22}e \\
	 \dot{\tau} &=& 1-\dz(\frac{\tau}{\rho}) \\
	 \end{array}\right. \hspace{2mm}  &
	 \;(x,e)\in \overline{\mathcal{C}} \mbox{ or }   0 \leq \tau \leq \Delta \\
	 \left\{\begin{array}{lll}
	 {x}^+ & = & x \\
	 {e}^+ & = & 0\\ 
	 {\tau}^+ & = & 0\\ 
	 \end{array} \right. \hspace{16mm} & 
         \;(x,e) \in \overline{\mathcal{D}} \mbox{ and }  \tau \geq \Delta \\
	 \begin{array}{lll}
		y = Cx  \hspace{16mm} & &  \\
	 \end{array} &
 \end{array}
\end{equation}
where
$F_{11} := (A+BC)$ is Hurwitz by Assumption \ref{assume:well-posed}, 
$F_{12} := B$,
$F_{21} :=-C(A+BC)$,
$F_{22} :=-CB$,
and the relation between flow sets and jump sets before and after the coordinate transformation 
is given by $\mathcal{C}_\Delta :=$
\begin{equation}
\{(x,\nu,\tau) \in \realn \!\times\! \real^q \!\times\! \real \,|\, 
(x,\nu-y)\in \overline{\mathcal{C}} \mbox{ or } 0 \leq \tau \leq \Delta \},
\end{equation}
and by $\mathcal{D}_\Delta :=$
\begin{equation}
\{(x,\nu,\tau) \in \realn \!\times\! \real^q \!\times\! \real \,|\, 
(x,\nu-y)\in \overline{\mathcal{D}} \mbox{ and } \tau \geq \Delta \}.
\end{equation}
In what follows we use $F:=\smallmat{ F_{11} & F_{12}  \\ F_{21} & F_{22}}$.  

{
By coordinate transformation, the closed-loop state is now
directly related to the error $e$ between the current value of the output $y$
and the actual value of the samples $\nu$ transmitted to the controller.
In fact, the synchronous transmission policy will require the transmission of
a new sample when a particular relation between state $x$ and error $e$ holds.
This behavior is modeled by the definition of 
the sets $\overline{\mathcal{C}}$ and $\overline{\mathcal{D}}$.
In particular, 
from \eqref{eq:elazy_sys}, the typical behavior of the 
t-lazy closed loop is to transmit a new sample
only if the pair $(x,e)$ satisfies the criterion 
$(x,e) \in \overline{\mathcal{D}}$ and if at least $\Delta$ 
units of time have elapsed from the last transmission.
}

\subsection{The synchronous transmission policy}

The { synchronous} transmission policy 
$(\overline{\mathcal{C}},\overline{\mathcal{D}},\Delta,\rho)$ 
is based on two main parameters $\gamma_e>0$ and $\gamma_x>0$.
{
\begin{itemize}
\item $\gamma_e>0$ guarantees a proportionality between the norm of the error $e$ 
and the current state $x$, to avoid an asymptotic growth to infinity of the
error $e$ while $x$ remains bounded. The conditions on $\gamma_e>0$ for
the stability of the closed loop are very mild: every $\gamma_e>0$
guarantees stability;
 \item $\gamma_x>0$ specifies a bound on
the decay rate of a given Lyapunov function $V$. This bound is 
directly connected to the frequency of the sample transmissions,
since a transmission is required when the desired
decrease on the Lyapunov function is not guaranteed anymore.
The connection between sample transmission and Lyapunov-based
conditions is given by the sets $\overline{\mathcal{C}}$ and $\overline{\mathcal{D}}$.
\end{itemize}
The Lyapunov function $V$ is suitably defined within condition \textbf{(S1)} below,
and it is used in condition \textbf{(S2)} below to characterize 
$\overline{\mathcal{C}}$ and $\overline{\mathcal{D}}$. 
Note that, by definition, $\overline{\mathcal{C}}\neq \emptyset$.
}

\noindent\textbf{(S1)}
{Take} $\gamma_x>0$, $Q=Q^T>0$, and 
$V:\realn\!\times\!\real^q \!\to\!\real_{\geq 0}$, 
\begin{equation}
 V(x,e) := \frac{1}{2} \mymatrix{c}{x \\ e}^T \mymatrix{cc}{P_1 & 0 \\ 0 & P_2} \mymatrix{c}{x \\ e}, \label{V_fullstate} 
\end{equation}
such that
\begin{itemize}
 \item $P:=\left[
\begin{smallmatrix}
P_1 & 0 \\ 
0 & P_2 \\
\end{smallmatrix}\right]$  is symmetric
and positive definite,
\item $F_{11}^TP_1+P_1F_{11} \leq -Q$,
\item $\gamma_x I< Q$.
\end{itemize}
\textbf{(S2)} For any $\gamma_e>0$, define
\begin{equation}
\label{CD_fullstate}
\begin{array}{ll}
 \overline{\mathcal{C}} &:= \setof{(x,e)\,| 
\langle\nabla V(x,\!e), F\smallmat{x \\ e}\rangle  
\leq -\gamma_x|x|^2 \mbox{ and } |e| \leq \gamma_e |x|} \\
 \overline{\mathcal{D}} &:= \setof{(x,e)\,| 
\langle\nabla V(x,\!e), F\smallmat{x \\ e}\rangle  
\geq -\gamma_x|x|^2 \mbox{ or } |e| \geq \gamma_e |x|}. \\
\end{array} 
\end{equation}
We can now state the main result of the section, whose proof
is provided in Section \ref{sec:proof_thm1}.
\begin{theorem} 
\label{thm:GAS_synchronous}
Under Assumption \ref{assume:well-posed}, 
consider a transmission policy 
$(\overline{\mathcal{C}},\overline{\mathcal{D}},\Delta,\rho)$
which satisfies (\textbf{S1}) and (\textbf{S2}). 
Then, there exists $\Delta>0$ and $\rho>\Delta $ (sufficiently small) 
such that  the compact set 
\begin{equation}
\label{eq:A_sync}
 \mathcal{A} := \{0\}\times \{0\} \times [0,2\rho]
\subset \realn\times\real^q\times\real
\end{equation}
is globally asymptotically stable for the 
t-lazy closed-loop system \eqref{eq:elazy_sys}. Moreover,
if $B$ in \eqref{eq:plant} is full column rank, then 
$\mathcal{A}$ is globally exponentially stable.
\end{theorem}

{
The reader will notice that the asymptotic stability of the set 
$\mathcal{A}$ in \eqref{eq:A_sync} entails asymptotic stability
of the equilibrium $x=0$, which is the exponentially stable 
equilibrium of the original closed loop system
(by construction $\overline{\mathcal{C}}\cup \overline{\mathcal{D}} 
= \realn\times\real^q\times\real_{\geq 0}$, therefore every solution
to \eqref{eq:elazy_sys} is complete).
Moreover, under the mild hypothesis of $B$ full column rank,
the exponential stability of the original closed-loop system
is preserved by the synchronous transmission policy.

Note that, under Assumption \ref{assume:well-posed}, 
it is straightforward to see that 
the inequalities in (\textbf{S1}) are feasible. In fact, 
for any given $\gamma_x$, 
there exists a matrix $Q=Q^T > 0$ such that
$\gamma_x I< Q$. Then, by
Assumption \ref{assume:well-posed}, there exists a matrix 
$P_1=P_1^T>0$ which satisfies the condition 
$F_{11}^TP_1+P_1F_{11} \leq -Q$.
Note that the inequality $\gamma_x I< Q$ guarantees that
$\overline{\mathcal{C}} \neq \emptyset$. To see this, take $e=0$, then
$\langle\nabla V(x,\!e), F\smallmat{x \\ e}\rangle 
\leq -x^T\!Qx  < -\gamma_x |x|^2 $

Since the average sample transmission frequency is connected to the 
decay rate of the function $V$, this frequency can be partially regulated 
by suitably choosing $P_1$, $P_2$, $\gamma_x$, and $\gamma_e$. In the typical 
scenario, the transmission of one sample resets the error $e$, which
typically increases during the flow interval after the sample transmission,  
weighted by $P_2$. Then, possibly, the boundary of 
the set $\overline{\mathcal{C}}$ is reached and a new sample is transmitted.
In particular, from the definition of $\overline{\mathcal{C}}$, for 
an initial condition $e=0$ and $x\neq 0$, smaller values for $P_2$ 
guarantee longer flow intervals. In fact, for 
smaller values of $P_2$, $(x,e)$ satisfies 
$\langle\nabla V(x,\!e), F\smallmat{x \\ e}\rangle  > -\gamma_x|x|^2$
for larger values of $e$. In the limit, that is, for $P_2=0$ 
and $\gamma_e=\infty$ (i.e. no bound on $e$),
$e$ may grow unbounded and $\mathcal{A}$ is not necessarily stable. But 
this is forbidden by condition \textbf{(S1)}.
}

{
\begin{remark}[\emph{Comparison with selected literature}] \\
The synchronous transmission policy is closely related to the
event-triggered control approaches 
\cite{Anta08,Anta10,Seuret11,Tabuada07,Wang08a} (which consider a more 
general class of nonlinear control systems) and \cite{Cervin08,Mazo08},
which propose transmission policies based on inequalities on the
measured output and on the state of the plant.
In particular, the transmission policy in \cite[Section IV]{Tabuada07} 
is similar to our synchronous policy, as shown by \cite[Equation (13)]{Tabuada07}
which also highlights some differences between our approach and the one in \cite{Tabuada07}. 
Specifically, for any choice of the event-triggering conditions used in \cite{Tabuada07} 
(which are expressed simply as inequalities involving the norms $|x|$ and $|e|$),
there always exist an equivalent choice of the sets $\overline{\mathcal{C}}$ 
and $\overline{\mathcal{D}}$ in \eqref{CD_fullstate} and of $\Delta$ in \eqref{eq:lazy_sys_timer}
(with $\Delta<\tau$ and $\tau$ given by \cite[Corollary IV.1]{Tabuada07}) yielding exactly the same events;
the converse is not true, in general, due to the restricted dependence from $|x|$ and $|e|$
(instead of $x$, $e$) in \cite{Tabuada07}. { See also the simulation results of Section~\ref{sec:example1} 
where we compare these two approaches.}
As for \cite{Seuret11}, it proposes two event-triggering strategies and generalizes the 
assumptions in \cite{Anta08,Anta10,Tabuada07,Wang08a}.
In particular, the proposed strategies are based on the  
relaxation of the input-to-state stability assumptions on the underlying (not triggered) system
to global asymptotic stability assumptions. Unfortunately, this generalizations
do not provide improvements on the transmission rate for LTI systems since 
on LTI systems the two properties are equivalent.
Both \cite{Cervin08} and \cite{Mazo08} focus on more transmission-related implementation issues,
thus being less related to the basic stabilization problem considered here;
in \cite{Cervin08} the optimization of a stochastic performance index 
measuring the state variance is considered 
for a network of simple dynamical systems, whereas in \cite{Mazo08}
the issue of balancing energy consumption in different nodes is tackled.
Finally, alternative approaches are simpler to implement but require more bandwidth,
like \cite{Carnevale07}, which guarantees transmissions to happen 
before the expiration of the maximum allowable delay compatible with stability preservation.

While the main novelty of this paper as compared to previous approaches is in the proposed asynchronous transmission policy,
even in the synchronous case we provide some advantages and novelties.
First, in general, global exponential stability is guaranteed with 
less conservative bounds with respect to the current literature
(see e.g. the above discussion about \cite{Tabuada07}).
Then, the formulation within the hybrid systems framework of \cite{GoebelCSM09,Goebel12} 
automatically provides some levels of robustness which are guaranteed by the framework itself, 
as well as several analysis tools that make it easier to establish some relevant properties
and additional results (see e.g. the output feedback results in Section~\ref{sec:output_feedback}).
Finally, an additional novelty is the introduction of a timer (the state $\tau$) 
within the sensors; while simple to implement, such modification 
enforces a minimum interval between consecutive transmissions of samples,
meanwhile preserving stability (more recently, the same idea was used in \cite{Seuret11}).
\end{remark}
}

\subsection{Proof of Theorem \ref{thm:GAS_synchronous}} 
\label{sec:proof_thm1}
The proof technique is inspired by \cite[Example 27]{GoebelCSM09}.
{

From the assumptions of the theorem, we provide a Lyapunov
function $W$ in \eqref{eq:W(X)}, we show that $W$ is non-increasing 
at each sample transmission (at jumps) in \eqref{eq:proofthm1_0},
and we show that $W$ is non-increasing during flows, by decomposing
the analysis in two parts: (i) for $\tau\in[0,\Delta]$, and
(ii) for $\tau\geq \Delta$ and $(x,e) \in \overline{\mathcal{C}}$.
Combining these results with the invariance principle for
Hybrid systems in \cite[Theorem 23]{GoebelCSM09} and \cite{SanfeliceTAC07},
we prove global asymptotic stability. Finally, based on a recent result
in \cite[Theorem 2]{TeelTAC13}, we strengthen the asymptotic
convergence to an exponential one, by using the hypothesis that $B$
is full column rank.
}

{ \underline{Lyapunov function:}}
using $X$ for the aggregate state $\smallmat{x^T & e^T &\tau}^T$,
consider the following Lyapunov function
\begin{equation}
\label{eq:W(X)}
 W(X) = x^TP_1x + \exp((2\rho-\tau)\lambda)e^TP_2e
\end{equation}
where $\lambda >0$ is selected later.
Using the definitions $\underline{\alpha} :=  \lambda_{\min}(P)$ 
and $\overline{\alpha} := \max\{\lambda_{\max}(P_1),
\exp(2\rho\lambda)\lambda_{\max}(P_2)\}$ 
we get
$\underline{\alpha}|X|_\mathcal{A}^2 \leq W(X) \leq \overline{\alpha}|X|_\mathcal{A}^2$
{ (radially unbounded)}.

{ \underline{Lyapunov function at jumps:}} we have that
\begin{equation}
\label{eq:proofthm1_0}
W(X^+) -W(X)= 
-\exp((2\rho-\tau)\lambda)e^TP_2e \leq -e^TP_2e
\end{equation}
for each $X$ such that $(x,e)\in\overline{\mathcal{D}} \mbox{ and } \tau \geq \Delta$. 

{ \underline{Lyapunov function on flows:}}
the analysis is developed by considering two cases:
(i)~$\tau\in[0,\Delta]$, and (ii)~$\tau \geq \Delta$.

For (i), using $\varphi(\rho,\lambda,\tau) := \exp((2\rho-\tau)\lambda)$, 
considering $1-\dz(\frac{\tau}{\rho})=1$, and defining
$\lambda_0 := \lambda_{\min}(Q)$, 
$\gamma_1 := 2|P_1F_{12}|$,
$\gamma_2 := 2|P_2F_{21}|$
$\gamma_3 := 2|P_2F_{22}|$, and
$\gamma_4 := \lambda_{\min}(P_2)$, 
we get
\begin{equation}
\label{eq:proofthm1_a}
\begin{array}{lll}
\dot{W} &\leq & -x^TQx + 2x^TP_1F_{12}e \\
& & + 2\varphi(\rho,\lambda,\tau)e^TP_2(F_{21}x + F_{22}e) \\
& &- \lambda\varphi(\rho,\lambda,\tau) e^TP_2e\\
&\leq &
-\lambda_0|x|^2 + \gamma_1|x||e| + \gamma_2\varphi(\rho,\lambda,\tau)|e||x| \\
& & + 
\gamma_3\varphi(\rho,\lambda,\tau)|e|^2   - \lambda \gamma_4\varphi(\rho,\lambda,\tau) |e|^2.
\end{array}
\end{equation}
Exploiting the inequality $ab\leq \frac{1}{\varepsilon}a^2 + \varepsilon b^2$, 
where $\varepsilon>0$, and $a,b\in\real$, 
from \eqref{eq:proofthm1_a} we get
\begin{equation}
\label{eq:proofthm1_b}
\begin{array}{lll}
\dot{W} 
&\leq &
-\lambda_0|x|^2 + \gamma_1\frac{\varepsilon}{\varphi(\rho,\lambda,\tau)}|x|^2 + 
\gamma_1\frac{\varphi(\rho,\lambda,\tau)}{\varepsilon}|e|^2 \\
& & + 
\gamma_2\varepsilon|x|^2 
+ \gamma_2\frac{\varphi(\rho,\lambda,\tau)^2}{\varepsilon}|e|^2 \\
& & + 
\gamma_3\varphi(\rho,\lambda,\tau)|e|^2   - \lambda \gamma_4\varphi(\rho,\lambda,\tau) |e|^2 \\
& \leq & 
(-\lambda_0 + \gamma_1 \varepsilon +
\gamma_2\varepsilon) |x|^2 \\
& & + 
\varphi(\rho,\lambda,\tau)(\underbrace{\frac{\gamma_1}{\varepsilon} +
\frac{\gamma_2\varphi(\rho,\lambda,0)}{\varepsilon} + \gamma_3 - \lambda \gamma_4}_{=:\lambda_1})
 |e|^2 \\
& \leq &{
-\frac{1}{2}\lambda_0|x|^2 -\varphi(\rho,\lambda,\tau)\lambda_1 |e|^2 }
\quad \forall \tau\in[0,\Delta],
\end{array}
\end{equation}
{where the first term of the last inequality follows from 
the selection of $\varepsilon := \frac{\lambda_0}{2(\gamma_1 + \gamma_2)}$,
while $\lambda_1> 0$ is achieved by picking
$\lambda,\rho$ such that
$\lambda > \frac{\gamma_3}{\gamma_4} 
+\frac{\gamma_1}{\gamma_4\varepsilon}
+\frac{\gamma_2\varphi(\rho,\lambda,0)}{\gamma_4\varepsilon}$,
which can always be satisfied by picking $\lambda>0$ sufficiently large and 
$\rho> \Delta > 0$ sufficiently small (i.e. for the design parameter
$\Delta$ sufficiently small). 
For instance, define $c_1 := \frac{\gamma_3}{\gamma_4} 
+ \frac{\gamma_1}{\gamma_4\varepsilon}$ and
$c_2 :=  \frac{\gamma_2}{\gamma_4\varepsilon}$,
the inequality above reads 
$\lambda  > c_1 + c_2 \varphi(\rho,\lambda,0)$,
which holds for 
$\lambda := c_1+2c_2$ and $\rho$ sufficiently
small, since $\varphi(\rho,\lambda,0) \to 1$ as $\rho  \to 0$.
}

For (ii), $\tau > \Delta$ implies $(x,e)\in\overline{\mathcal{C}}$.
Thus, considering $1-\dz(\frac{\tau}{\rho})\geq 0$, and
using $\varphi(\rho,\lambda,\tau) := \exp((2\rho-\tau)\lambda)$, 
we get
\begin{eqnarray}
\label{eq:proofthm1_c}
\dot{W} &\leq & -x^T(F_{11}^T P_1 + P_1F_{11})x + 2x^TP_1F_{12}e \\ \nonumber
& &+\, 2\varphi(\rho,\lambda,\tau)e^TP_2(F_{21}x + F_{22}e) \\ \nonumber
&= & 
 \langle\nabla V(x,\!e), F\smallmat{x \\ e}\rangle\\ \nonumber
& & +\, 2(\varphi(\rho,\lambda,\tau)-1)e^TP_2(F_{21}x + F_{22}e) \\ \nonumber
&\leq & -\gamma_x |x|^2
 + 2(\varphi(\rho,\lambda,\tau)-1)e^TP_2(F_{21}x + F_{22}e) \\ \nonumber
&\leq & -\gamma_x |x|^2
 + 2(\varphi(\rho,\lambda,\tau)-1)(\gamma_2|e||x| + \gamma_3|e|^2) \\ \nonumber
&\leq & -\gamma_x |x|^2
 + 2(\varphi(\rho,\lambda,\Delta)-1)(\gamma_2\gamma_e|x|^2 + \gamma_3\gamma_e^2|x|^2) \\ \nonumber
&\leq & -\frac{\gamma_x}{2} |x|^2 \ ,
\end{eqnarray}
{
where the last inequality holds for $\rho> \Delta$ sufficiently small.
For example, take $\rho = 2\Delta$, then
$\varphi(\rho,\lambda,\Delta) = \varphi(2\Delta,\lambda,\Delta)
= \exp(3\Delta\lambda) \to 1$, as $\Delta \to 0$.
As in \eqref{eq:proofthm1_b}, the decreasing of $W$ in
\eqref{eq:proofthm1_c} is achieved by selecting the design parameter
$\Delta$ sufficiently small. 
}

{
Both cases (i) and (ii) are then covered by considering 
$\rho \in(0, \min\{\frac{\ln(c_3/c_2)}{2(c_1+c_3)},
\frac{1}{2\lambda}
\log(\frac{\gamma_x}{4(\gamma_2\gamma_e + \gamma_3\gamma_e^2)} +1) 
+ \frac{\Delta}{2}\}).$
}

{ \underline{GAS of the set $\mathcal{A}$ by invariance principle:}}
since the t-lazy closed-loop system in \eqref{eq:elazy_sys} 
satisfies the basic conditions of \cite{GoebelCSM09} (see Section~\ref{sec:preliminaries}), 
from the inequalities above, following \cite[Theorem 23]{GoebelCSM09} 
or \cite{SanfeliceTAC07} $\mathcal{A}$ is stable. Moreover, for any given $\mu>0$,
consider the level curve given by $\ell(\mu) = \{X\,|\,\,W(X)=\mu\}$.
Suppose now
that $X(0,0)\in \ell(\mu)\cap\overline{\mathcal{D}}$ with $e(0,0)\neq 0$.
Then, from \eqref{eq:proofthm1_0} $W(X)$ decreases. Thus, 
suppose $X(0,0)\in \ell(\mu)\cap\overline{\mathcal{D}}$ with $e(0,0)=0$.
From the definition of $\overline{\mathcal{D}}$ in \eqref{CD_fullstate},
necessarily $x=0$, thus $W(X)=0$ (in fact, for $e=0$ and $x\neq 0$, 
$X\notin \overline{\mathcal{D}}$).
During flows each solution $X$ such that 
$X(0,0)\in \ell(\mu)\cap\overline{\mathcal{C}}$ and $x(0,0)\neq 0$  
guarantees that $W(X)$ decreases 
(by \eqref{eq:proofthm1_b} and \eqref{eq:proofthm1_c}).  
For $X\in \ell(\mu)\cap \overline{\mathcal{C}}$ and $x(0,0)=0$, 
considering $\tau \leq \Delta$, we have that $W$ decreases 
(by \eqref{eq:proofthm1_b}), while considering $\tau > \Delta$
we have $(x,e)\in \overline{\mathcal{C}}$ which implies 
$|e| \leq \gamma_e|x| \leq 0$, thus $W(X)=0$.
Thus, using the fact that $W(X)$ is radially unbounded and
no complete solutions remain within $\ell(\mu)$, 
by \cite[Theorem 23]{GoebelCSM09}, the set $\mathcal{A}$ is globally asymptotically stable.

{ \underline{Exponential stability of the set $\mathcal{A}$:}}
it follows from the application of 
\cite[Theorem 2]{TeelTAC13}. 
For instance, 
decompose the state of the t-lazy closed-loop system 
in $\xi_1 = (x,e)$ and $\xi_2 = \tau$. Then,
conditions 1)-3) of \cite[Assumption 1]{TeelTAC13}
are satisfied. 
Moreover, $B$ full column rank implies 
the observability of the pair $(\smallmat{I_n & 0},F)$. In fact, 
using the linear transformation $T:=\smallmat{I & 0 \\ -C & I}$,
we have $T^{-1} = \smallmat{I & 0 \\ C & I}$ from which
$ \smallmat{A & B \\ 0 & 0} = T^{-1} F T$ and $\smallmat{I_n & 0} = \smallmat{I_n & 0} T$. 
Thus, the observability of the pair $(\smallmat{I_n & 0},F)$
can be established via observability PBH test on the pair 
$(\smallmat{I_n & 0},\smallmat{A & B \\ 0 & 0})$, that is, 
\begin{equation}
\label{eq:rank}
\mathrm{rank}\mymatrix{c|c}{A -sI & B \\ \hline  0 & -sI \\ \hline I_n & 0} =n+q \quad \forall s\in \mathbb{C}, 
\end{equation}
which holds when $B$ is full column rank.
Therefore, combining the observability of $(\smallmat{I_n & 0},F)$
with \eqref{eq:proofthm1_0}-\eqref{eq:proofthm1_c} 
and with the bound on $W$ given after \eqref{eq:W(X)}, 
condition 4) of \cite[Assumption 1]{TeelTAC13} is satisfied.
Finally, the jumps of the t-lazy closed-loop 
system satisfy an average dwell-time constraint, since for
each solution $X$ we have that 
$(t,j)\in \dom\,X$ implies $j \leq \frac{t}{\Delta}$,
which satisfies condition 5) of 
\cite[Assumption 1]{TeelTAC13}.
Thus, from \cite[Theorem 2]{TeelTAC13}, $\mathcal{A}$
is globally exponentially stable.
\hfill \hspace*{1pt} \hfill $\square$

{
\begin{remark}
The use of the timer $\tau$ guarantees a minimum dwell time $\Delta>0$ 
between transmissions. 
The proof of Theorem~\ref{thm:GAS_synchronous} 
provides a conservative bound on $\Delta$,
which leads to excessively small values for $\Delta$,
as revealed by the simulations. Relaxations on the bounds on 
$\Delta$ are possible but we will not pursue this analysis here.
Note that there is a particular
initial conditions from which a transmission may occur every $\Delta$ times,
like, for example, $x=0$ and $e=0$. But such a transmission is only apparent,
since $e^+ = e = 0$, thus the new transmission may be neglected.
\end{remark}
}

\section{State feedback: asynchronous transmission}
\label{sec:asynchronous}

\subsection{The error dynamics}

\begin{figure}[ht!]
\centering
\includegraphics[width=0.6\columnwidth]{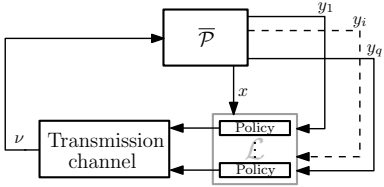}
 	\caption{The transmission-lazy closed-loop system under the asynchronous transmission policy.}
 	\label{fig:SLa}
\end{figure}

We consider an asynchronous transmission policy
in which each sensor autonomously decides 
whether or not to transmit a new sample,
based on its own state $\nu_i$, the timer $\tau_i$, 
and the state $x$ (which is assumed to be available to all sensors).
{ We call \emph{asynchronous} such a policy to 
underline the fact that the sensors transmit their measurements
at independent times, and at different rates. 
The decision of a sample transmission is \emph{autonomous} for each sensor in the 
precise sense that the single sensor does not need to know the error
of any other sensor to decide its transmission. However, it uses the
information on the state of the plant. A block diagram representing this scheme is
shown in Figure~\ref{fig:SLa}.}

{
Consider the system in \eqref{eq:lazy_sys} with each
sensor equipped with its own timer: $p=q$ and $\tau\in\real^p$. 
As in the previous section, the asynchronous transmission depends on
Lyapunov-like conditions used to construct the sets 
$\mathcal{C}_i,\mathcal{D}_i \subseteq \realn\times\real\times\real$, $i\in\{1,\dots,q\}$.
Each pair $\mathcal{C}_i,\mathcal{D}_i$
specifies when the $i$th sensor may transmit
a new sample. 

To allow for an asynchronous update of the measurement vector, we define}
the functions $g$ and $h$  as follows
\begin{equation}
\label{eq:g_async}
g(x,\nu,\tau):=
\smallmat{\eta_1 \\ \vdots \\ \eta_q}
\mbox{ where } \\
\left\{\begin{array}{ll}
 \eta_i = (Cx)_i & \mbox{ if }(x,\nu_i,\tau_i)\in {\mathcal{D}}_i \\
 \eta_i = \nu_i & \mbox{ otherwise,}
\end{array} \right.
\end{equation}
\begin{equation}
\label{eq:h_async}
h(x,\nu,\tau):=
\smallmat{\eta_1 \\ \vdots \\ \eta_q}
\mbox{ where } \\
\left\{\begin{array}{ll}
 \eta_i = 0 & \mbox{ if }(x,\nu_i,\tau_i)\in {\mathcal{D}}_i \\
 \eta_i = \tau_i & \mbox{ otherwise,}
\end{array} \right.
\end{equation}
where $(Cx)_i,\nu_i,\tau_i$ are the $i$th elements of the vectors $Cx,\nu,\tau$, 
respectively. 
Finally, the asynchronous transmission model can be completed 
 by defining the sets $\mathcal{C}_\Delta$ and $\mathcal{D}_\Delta$ 
in \eqref{eq:lazy_sys} as follows
\begin{equation}
\label{CD_fullstate3}
\begin{array}{lll}
 {\mathcal{C}_\Delta} &=& \{(x,\nu,\tau)\,|\, 
\forall i\; (x,\nu_i,\tau_i)\in {\mathcal{C}}_i \} \\
 {\mathcal{D}_\Delta} &=& \{(x,\nu,\tau)\,|\, 
\exists i\; 
 (x,\nu_i,\tau_i)\in {\mathcal{D}}_i \}. \\
\end{array}
\end{equation}
Thus, the definition of an asynchronous transmission policy 
$(\mathcal{C}_\Delta,\mathcal{D}_\Delta,\Delta,\rho)$ is equivalent to the definition of
$(\mathcal{C}_1,\dots,\mathcal{C}_q,\mathcal{D}_1,\dots,\mathcal{D}_q,\Delta,\rho)$.

{
The asynchronous transmission mechanism modeled by  
\eqref{eq:g_async}, \eqref{eq:h_async} \eqref{CD_fullstate3} 
can be easily understood by considering the following scenario.
}
Suppose that $(x,\nu_i,\tau_i)$ belongs to $\mathcal{D}_i$
for some $i$, and $(x,\nu_j,\tau_j) \notin \mathcal{D}_i$ for $j\neq i$. 
Looking at \eqref{CD_fullstate3}, the t-lazy closed-loop system 
$\mathcal{S}_\mathcal{L}$ may jump. Then,
from the definition of $g$ and $h$  in \eqref{eq:g_async}, 
\eqref{eq:h_async}, only the $i$th sample
will be transmitted, since $\nu_i^+ = y_i$, and $\tau_i^+=0$, 
while $\nu_j^+=\nu_j$ and $\tau_j^+=\tau_j$.

Following the approach of previous section, we present the asynchronous
transmission policy by using the coordinate 
transformation $(x,e)=(x,\nu-y)$, from which \eqref{eq:lazy_sys} becomes
\begin{equation}
\label{eq:elazy_sys2}
\begin{array}{ll}
	 \left\{ \begin{array}{lll}
	 \dot{x} & = & F_{11}x + F_{12}e  \\
  	 \dot{e} & = & F_{21}x  +F_{22}e \\
	 \dot{\tau} &=& \overline{1}-\dz(\frac{\tau}{\rho}) \\
	 \end{array}\right. \hspace{2mm}  & 
	 \forall i\,\big(
	  (x,e_i)\!\in\! \overline{\mathcal{C}}_i 
	  \mbox{ or } 0 \!\leq\! \tau_i \!\leq\! \Delta
	  \big) \\
	 \left\{\begin{array}{lll}
	 {x}^+ & = & x \\
	 {e}^+ & = & g(x,e\!+\!y,\tau) - y \\ 
	 {\tau}^+ & = & h(x,e\!+\!y,\tau)\\ 
	 \end{array} \right.  & 
         \exists i\,\big(
	  (x,e_i)\!\in\! \overline{\mathcal{D}}_i 
	  \mbox{ and } \tau_i \!\geq\! \Delta
	  \big)  \\
	 \begin{array}{lll}
		y = Cx  \hspace{16mm} & & \\
	 \end{array} &
 \end{array}
\end{equation}
from which, for each $i$, the sets $\overline{\mathcal{C}}_i,\overline{\mathcal{D}}_i$
are connected to the sets $\mathcal{C}_i,\mathcal{D}_i$ by the following relation
$\mathcal{C}_i :=
\{(x,\nu_i,\tau_i) \in \realn \times \real \times \real \,|\, 
(x,\nu_i-y_i)\in \overline{\mathcal{C}}_i \mbox{ or } 0 \leq \tau_i \leq \Delta \}$
and 
$\mathcal{D}_i :=
\{(x,\nu_i,\tau_i) \in \realn \times \real \times \real \,|\, 
(x,\nu_i-y_i)\in \overline{\mathcal{D}}_i \mbox{ or } \tau_i \geq \Delta \}$

{
Summarizing, from \eqref{eq:elazy_sys2} and from the definition of 
$\overline{\mathcal{C}}_i$ and $\overline{\mathcal{D}}_i$,
if $(x_i,e_i)\in \overline{\mathcal{D}}_i$ and $\tau_i \!\geq\! \Delta$,
then the jump map of \eqref{eq:elazy_sys2} guarantees that} 
$e_i^+=0$ and $\tau_i^+=0$, otherwise $e_i^+ = e_i$ and $\tau_i^+=\tau_i$. 
Moreover, the intersample time between two consecutive resets of each sensor $i$ 
is greater than or equal to $\Delta$, since, for each sensor $i$, 
resets are enabled only if the internal timer $\tau_i \geq \Delta$.

{
\begin{remark}
The transmissions of two or more sensors at the same time is
modeled by a sequence of two or more consecutive resets. This
case may
occur when  two or more indices $i$ satisfy the existential quantifier in \eqref{CD_fullstate3}.
In such a case the jump rule is given by the union of two or more
update laws in \eqref{eq:g_async}, \eqref{eq:h_async}. This definition produces
an outer semicontinuous set-valued jump map, thereby guaranteeing robustness 
(see \cite{Goebel06}). We do not elaborate further on robustness in this section,
postponing the analysis of the robustness of the proposed algorithms
to Section \ref{sec:robustness}.
\end{remark}
}
\subsection{The asynchronous transmission policy}

The { asynchronous} transmission policy is based on three parameters,
$\gamma_x > 0$, $\epsilon\in[0,\frac{1}{p}]$ and $\alpha\in\real^q$. 
{
\begin{itemize}
\item $\gamma_x$ is used to establish a bound on the decay rate of a suitable Lyapunov function.
This parallels the role of $\gamma_x$ in the previous section.
\item The constant $\epsilon \in [0,\frac{1}{p}]$ is a lower bound on the value of each
element of the vector $\alpha$. The presence of a lower bound allows for the
possibility of varying the gains $\alpha$ at runtime (for performance improvement) without losing closed-loop stability. 
\item Each element $\alpha_i$ of $\alpha$ satisfies
the condition $\alpha_i>\varepsilon$. Moreover, $\sum_{i=1}^q \alpha_i = 1$
and each $\alpha_i$ is a weight on the achievable decay rate associated to sensor $i$.
For example,  when $\alpha_i$ is small the $i-$th sensors transmits at higher
frequency. Therefore, the ratio between different elements of the vector $\alpha$ is
in direct relation to the transmission rate of each sensor.
\end{itemize}
The conditions on the Lyapunov function are given in \textbf{(A1)} below. 
Flow and jump sets are based on this Lyapunov function and are given in \textbf{(A2)}.
Based on these parameters, we formulate the 
asynchronous transmission
policy as follows.
}

\noindent\textbf{(A1)} 
Take $\gamma_x>0$, $\epsilon\in[0,\frac{1}{q}]$, $Q=Q^T>0$, and 
$V:\realn\!\times\!\real^q \!\to\!\real_{\geq 0}$, 
\begin{equation}
 V(x,e) := \frac{1}{2} \mymatrix{c}{x \\ e}^T \mymatrix{cc}{P_1 & 0 \\ 0 & P_2} \mymatrix{c}{x \\ e}, \label{V_fullstate2} 
\end{equation}
such that $P_1\!=\!P_1^T\!>0$, $P_2=\diag(p_1,\dots,p_q)>0$ and 
\begin{itemize}
\item $F_{11}^TP_1+P_1F_{11} \leq -Q$,
\item $ \frac{\gamma_x}{\epsilon} I< Q$.
\end{itemize}

\noindent\textbf{(A2)} Define 
\begin{equation}
a  := 2|P_1 F_{12}|,\quad
b := 2 |F_{21}|,\quad 
c:=  2 |F_{22}|,
\end{equation}  
and for each 
for each $i\in\{1,\dots,q\}$, define  $\overline{\mathcal{C}}_i$ and  
$\overline{\mathcal{D}}_i$ respectively as 
\begin{equation}
\label{CDi_fullstate2}
\begin{array}{lll}
\{(x,e_i)|
-\alpha_ix^T\!Qx \!+\!  (a\!+\!bp_i) |x||e_i| \!+\! cp_ie_i^2 & \leq & -\gamma_x|x|^2 \}\\
\{(x,e_i)|
-\alpha_ix^T\!Qx \!+\!  (a\!+\!bp_i) |x||e_i| \!+\! cp_ie_i^2 & \geq & -\gamma_x|x|^2 \}.\\
\end{array}
\end{equation}
We can now provide the main result of the section, 
whose proof is provided in Section \ref{sec:proof_thm1}. 
\begin{theorem} 
\label{thm:GAS_asynchronous}
Under Assumption \ref{assume:well-posed}, consider a transmission policy 
$(\overline{\mathcal{C}}_1 , \dots , \overline{\mathcal{C}}_q$,
$ \overline{\mathcal{D}}_1 , \dots,\overline{\mathcal{D}}_q , \Delta,\rho)$
which satisfies (\textbf{A1}) and (\textbf{A2}). 
Then, if $\sum_{i=0}^q \alpha_i=1$ and each 
$\alpha_i>\epsilon$, there exists $0<\Delta<\rho$ sufficiently small
such that the compact set
\begin{equation}
\label{eq:A_async} 
\mathcal{A} := \{0\}\times \{0\} \times [0,2\rho]^q
\subset \realn\times\real^q\times\real^q 
\end{equation}
is globally asymptotically stable for the 
t-lazy closed-loop system \eqref{eq:elazy_sys2}. 
Moreover,
if $B$ in \eqref{eq:plant} is full column rank, then 
$\mathcal{A}$ is globally exponentially stable.
\end{theorem}

{
Theorem~\ref{thm:GAS_asynchronous} establishes asymptotic
stability of the set $\mathcal{A}$ in the asynchronous
case, paralleling the synchronous results of Theorem~\ref{thm:GAS_synchronous}.
Like in the synchronous case,
all maximal solutions to \eqref{eq:elazy_sys2} are complete.
It is worth to note
however that the asynchronous transmission policy does not subsume the 
synchronous transmission policy. To see this, consider the implementation
of both approaches to a single input single output (SISO) system. 
In this scenario conditions \eqref{CDi_fullstate2} 
are more conservative than conditions \eqref{CD_fullstate} due to the
presence of extra terms that may
lead to a higher sample transmissions rate. In fact, 
\eqref{CDi_fullstate2} are designed to (conservatively) compensate for the presence of 
many t-lazy sensors, which are not present in the SISO case.
}

For $i=1,\dots,q$, $p_i$ can be considered as a weight on the 
single sensor error, and the combination of $\alpha_i$ and $p_i$ 
can be used to increase the update-rate 
of one sensor with respect to the others.
For example, considering each $p_i=1$, 
a greater $\alpha_i$ allows for a larger error 
bound on the $i$th sensor, thus the update-rate of that sensor 
decreases. 
Note also that each $\alpha_i$ can be modified at runtime. 
As long as $\sum_{i=0}^q \alpha_i=1$ and each 
$\alpha_i>\varepsilon>0$, global asymptotic stability is preserved. 

As shown in previous section, 
$\overline{\mathcal{C}}_i \neq \emptyset$. To see this,
consider $e=0$, then 
$-\alpha_i x^TQx \leq -\epsilon x^T Q x \leq \gamma_x|x|^2$,
from the last condition in \textbf{(A1)}. 
Looking at the definitions of $\overline{\mathcal{C}}_i$ and $\overline{\mathcal{D}}_i$,
note also that the information on the full state vector $x$ 
can be replaced by $|x|$ and $|x^TQx|$ only { which reduces 
dramatically the quantity of information required by each sensor}. 
Moreover, 
the result of Theorem \ref{thm:GAS_asynchronous}
still holds if $x^T Q x$ in \eqref{CDi_fullstate2} is 
replaced by $\lambda_{\min}(Q)|x|^2$ and in this case each t-lazy sensor
may decide its transmission by using only $e_i$ and $|x|$. 
Clearly, higher transmission frequency may occur 
since conservativeness is introduced.

\begin{remark}[{\emph{Comparison with selected literature}}] \\
An asynchronous transmission policy for a closed-loop system
defined by the interconnection of several linear systems can be found in \cite{Wang08b}, 
where a separated triggering condition for each system is provided and,
under specific decoupling conditions, it guarantees the stability of the 
interconnected system. 
{ 
As compared to that approach,
 our asynchronous transmission policy 
does not require any decoupling condition 
at the cost of using an additional information on the state of the 
controller-plant cascade, which is 
shared among the sensors.
This shared information is used to decide whether
or not to transmit a sampled output measurement $y_i$, without requiring the
transmission of the full output vector $y$.}
 A complementary approach can be found in \cite{Tabuada10} where
each sensor may decide to trigger a transmission of the whole vector $y$, 
based on its local error $e_i$ and its partial knowledge of the state vector $x$.
\end{remark}

{
\subsection{Proof of Theorem \ref{thm:GAS_asynchronous}}
\label{sec:proof_thm2}

The proof of the theorem follows the line of the proof of Theorem \ref{thm:GAS_synchronous}.
For instance, we introduce a Lyapunov function and we show the nonincresing features
of the function
at jumps and during flows. Then, based on the established inequalities, we apply
the invariant principle in \cite[Theorem 23]{GoebelCSM09} to show global asymptotic
stability of the set $\mathcal{A}$. Global exponential stability is then established
by invoking \cite[Theorem 2]{TeelTAC13}, under the mild assumption that
the matrix $B$ is full-column rank.
}

{ \underline{Lyapunov function:}} 
using $X$ for the aggregate state $\smallmat{x^T & e^T &\tau}^T$,
consider the Lyapunov function 
$W:\realn\times\real^q\times\real \to \real_{\geq 0}$ given by
\begin{equation}
\label{eq:proofthm2_0}
 W(X) = x^TP_1x + \sum_{i=1}^q p_i\exp( (2\rho-\tau_i)\lambda ) e_i^2
\end{equation}
where $\lambda>0$. Then, using $\underline{\alpha}$, $\overline{\alpha}$
defined in the proof of Theorem \ref{thm:GAS_synchronous}, 
$\underline{\alpha}|X|_\mathcal{A}^2 \leq W(X) \leq \overline{\alpha}|X|_\mathcal{A}^2$.

{ \underline{Lyapunov function at jumps:}} we have that
\begin{equation}
\label{eq:proofthm2_a}
W(X^+) -W(X)= -p_i\exp((2\rho-\tau_i)\lambda)e_i^2 
\end{equation}
for each $(x,e_i)\in\overline{\mathcal{D}}_i$ and $\tau_i \geq \Delta_i$,
and $i\in\{1,\dots, q\}$.

{ \underline{Lyapunov function on flows:}}
{ we first establish a convenient bound on the dynamics.} Using
$\varphi_i(\rho,\lambda,\tau_i) := \exp((2\rho-\tau_i)\lambda)$,
for $i\in\{1,\dots, q\}$, and 
$\varphi(\rho,\lambda,\tau) := 
\diag\{\exp((2\rho-\tau_1)\lambda),\dots,\exp((2\rho-\tau_q)\lambda)\}$,
the derivative of $W$ is bounded by 
\begin{equation}
\label{eq:proofthm2_b}
\begin{array}{lll}
\dot{W} &\leq & 
-x^TQx + 2x^T P_1F_{12} e  \\
& & + 2e^T\varphi(\rho,\lambda,\tau)P_2(F_{21}x + F_{22}e) \\
& & -\lambda \diag\left( 1-\dz(\frac{\tau}{\rho}) \right) 
e^T\varphi(\rho,\lambda,\tau)P_2e^T \\
&\leq & 
-x^T\!Qx + a|x||e|  
 + \sum\nolimits\limits_{i=1}^q \!
\varphi_i(\rho,\lambda,\tau_i)p_i|e_i| (b|x| \!+\! c|e|) \\
& & -\lambda \sum\nolimits\limits_{i=1}^q 
\left( 1-\dz(\frac{\tau_i}{\rho}) \right) 
\varphi_i(\rho,\lambda,\tau_i)
p_i|e_i|^2 \\
&\leq & 
\sum\nolimits\limits_{i=1}^q 
-\alpha_i x^TQx + \Big(a+\varphi_i(\rho,\lambda,\tau_i)p_ib\Big)|x||e_i|  \\
& & + 
\varphi_i(\rho,\lambda,\tau_i)p_i 
\Big(c- \lambda ( 1-\dz(\frac{\tau_i}{\rho}))\Big) |e_i|^2. \vspace{-5mm}\\
\end{array} \vspace{1mm}
\end{equation}

{Then, we develop the analysis of \eqref{eq:proofthm2_b}
by considering two cases}. Using 
$\pi_i := 
-\alpha_i x^TQx 
+ \Big(a+\varphi_i(\rho,\lambda,\tau_i)p_ib\Big)|x||e_i|  
+ \varphi_i(\rho,\lambda,\tau_i)p_i 
\Big(c- \lambda ( 1-\dz(\frac{\tau_i}{\rho}))\Big) |e_i|^2 $
to simplify the notation, for each $i$, 
let us consider two cases: $(i)$~$\tau_i\in [0,\Delta]$
and $(ii)$~$\tau_i\geq\Delta$. 

For $(i)$, 
considering  $1-\dz(\frac{\tau_i}{\rho})=1$ and using the inequality
$s_1s_2 \leq \frac{1}{\varepsilon}s_1^2 + \varepsilon s_2^2 \ 
\forall s_1,s_2\in\real$, $\varepsilon>0$, we get
\begin{equation}
\label{eq:proofthm2_c}
\begin{array}{lll}
\pi_i 
&\leq& 
-\epsilon x^TQx 
+ \Big(a+\varphi_i(\rho,\lambda,\tau_i)p_ib\Big)|x||e_i|  \\
& &+ p_i\varphi_i(\rho,\lambda,\tau_i) 
(c- \lambda ) |e_i|^2 \\
&\leq&
-\epsilon\frac{\lambda_0}{2}|x|^2 -\lambda_1 \varphi_i(\rho,\lambda,\tau_i) |e_i|^2\
\end{array} 
\end{equation}
where the last inequality holds for $\lambda_0= \lambda_{\min}(Q)$,
and for some $\lambda,\rho,\lambda_1>0$ ($\rho$, $\lambda_1$ sufficiently small),
by using an argument similar to
\eqref{eq:proofthm1_a} and \eqref{eq:proofthm1_b}.
In fact, the right-hand side of the first inequality in 
\eqref{eq:proofthm2_c} is very similar to 
the right-hand side of the last inequality in \eqref{eq:proofthm1_a}.

For $(ii)$, since $\tau_i\geq \Delta$, we have that
$(x,e_i)$ belongs to $\overline{\mathcal{C}}_i$. Thus,
as a first step, we claim the existence of a bound 
$|e_i|\leq \gamma_i|x|$ for some 
$\gamma_i:= \frac{\varepsilon \lambda_{\min}(Q)}{c p_i}$, 
which follows from 
the definition of $\overline{\mathcal{C}}_i$ in 
\eqref{CDi_fullstate2}, by
\begin{equation}
\label{eq:proofthm2_d}
\begin{array}{l}
 -\! \alpha_ix^TQx \!+\!  (a\!+\!bp_i) |x||e_i| \!+\! cp_ie_i^2 \leq -\gamma_x|x|^2 \\
\Rightarrow  \quad
cp_i e_i^2  
\!\leq\! \lambda_{\max}(Q)|x|^2 \\
\Rightarrow  \quad
e_i^2 \leq \frac{ \lambda_{\max}(Q)}{c p_i} |x|^2.
\end{array}
\end{equation}
Then, as a second step, since $1-\dz(\frac{\tau_i}{\rho})\geq 0$
for $\tau_i\geq \Delta$, 
using the definition of $\overline{\mathcal{C}}_i$, we get 
\begin{equation}
\begin{array}{lll}
\label{eq:proofthm2_e}
\pi_i 
&\leq& 
-\alpha_i x^TQx + \Big(a+\varphi_i(\rho,\lambda,\tau_i)p_ib\Big)|x||e_i|  \\
& & + \varphi_i(\rho,\lambda,\tau_i)p_i c |e_i|^2 \\
&=& 
-\alpha_i x^TQx + (a+p_ib)|x||e_i| + p_i c|e_i|^2  \\
& & 
+\Big(\varphi_i(\rho,\lambda,\tau_i)-1\Big)p_i (b|x||e_i| + c|e_i|^2) \\
&\leq & -\gamma_x|x|^2 
+\Big(\varphi_i(\rho,\lambda,\Delta)-1\Big)p_i (b\gamma_i + c\gamma_i^2)|x|^2  \\
&\leq & -\frac{\gamma_x}{2}|x|^2 \vspace{-4mm}
\end{array} \vspace{1mm}
\end{equation}
where the last inequality holds for $\rho>0$ sufficiently small, 
as shown in \eqref{eq:proofthm1_c} for a similar setup.

Define now $\mathcal{I}_{\Delta}:=\{i\,|\, \tau_i \leq \Delta\}\subseteq \{1,\dots,q\}$
and use $|\mathcal{I}_{\Delta}|$ to denote the number of elements of 
$\mathcal{I}_{\Delta}$.
Then, for $0<\Delta<\rho$ and $\rho$ sufficiently small,  
from $(i)$, $(ii)$ and \eqref{eq:proofthm2_b} we get 
\begin{equation}
\label{eq:proofthm2_f}
 \dot{W}(X) 
\leq -q \min\!\left(\frac{\epsilon\lambda_0}{2},\frac{\gamma_x}{2} \right) |x|^2
- \sum\nolimits\limits_{i\in I_{\Delta}} \lambda_1\varphi_i(\rho,\lambda,\Delta) |e_i|^2. \\
\end{equation}
$\forall (x,e,\tau)$ such that 
$\forall i\big( (x,e_i)\!\in\! \overline{\mathcal{C}}_i 
	  \mbox{ or } 0 \!\leq\! \tau_i \!\leq\! \Delta \big)$.

{ \underline{GAS of the set $\mathcal{A}$ by invariance principle:}}
using the fact that \eqref{eq:elazy_sys2} 
satisfies the basic conditions of \cite{GoebelCSM09} (see Section~\ref{sec:preliminaries}), 
combining \eqref{eq:proofthm2_a}, \eqref{eq:proofthm2_f}, and 
the bounds on $W(X)$ defined after 
\eqref{eq:proofthm2_0}, stability follows from \cite[Theorem 23]{GoebelCSM09}.
To establish global asymptotic stability (GAS) we proceed as 
in the proof of Theorem~\ref{thm:GAS_synchronous}.
For any given $\mu>0$,
consider the level curve given by 
$\ell(\mu) = \{X\,|\,\,W(X)=\mu\}$.
Suppose now that $X(0,0)\in \ell(\mu)$. From
\eqref{eq:proofthm2_f}, 
each solution $X$ from $\tau(0,0) \leq \Delta$ or
from $(x(0,0),e_i(0,0))\in\overline{\mathcal{C}_i},x(0,0)\neq 0$
guarantees that $W$ decreases. Moreover, each solution from 
$(x(0,0),e_i(0,0))\in\overline{\mathcal{C}_i}$, $x(0,0)= 0$, 
$\tau> \Delta$ necessarily has $e(0,0)=0$, 
which follows from \eqref{eq:proofthm2_d}.
Suppose now $X(0,0)\in \ell(\mu)$ and 
$(x,e_i)\in \overline{\mathcal{D}}_i$
for some $i$.  
From \eqref{eq:proofthm2_a},
each solution $X$ from 
$e_i(0,0)\neq 0$ guarantees that $W(X)$ decreases. 
If $e_i(0,0) = 0$, from \textbf{(A2)}, $\frac{\gamma_x}{\epsilon} < Q$,
and the definition of $\overline{\mathcal{D}}_i$ in \eqref{CDi_fullstate2},
necessarily $x=0$.
Thus, using the fact that $W(X)$ is radially unbounded and
no complete solutions remain within $\ell(\mu)$, 
by \cite[Theorem 23]{GoebelCSM09}, 
the set $\mathcal{A}$ is GAS.

{ \underline{Exponential stability of the set $\mathcal{A}$:}}
it can be 
established by following an argument similar to 
to the proof of Theorem \ref{thm:GAS_synchronous}.
For instance, using $\xi_1 = (x,e)$ and $\xi_2 = \tau$,
conditions 1)-3) of \cite[Assumption 1]{TeelTAC13}
are satisfied. $B$ full column rank implies the observability
of $(\smallmat{I_n & 0},F)$, which combined with
\eqref{eq:proofthm2_a}, \eqref{eq:proofthm2_f} and with the bound on
$W$ given after \eqref{eq:proofthm2_0}, 
condition 4) of \cite[Assumption 1]{TeelTAC13} 
is satisfied. Finally, the jumps of the t-lazy closed-loop 
system satisfy an average dwell-time constraint, since 
for each solution $X$, we have that
$(t,j)\in \dom\,X$ implies $j \leq q\frac{t}{\Delta}$,
(each sensor may reset at most  $\frac{t}{\Delta}$ times).
Thus, from \cite[Theorem 2]{TeelTAC13}, 
$\mathcal{A}$ is globally exponentially stable.
\hfill \hspace*{1pt} \hfill $\square$

\section{Output feedback approach}
\label{sec:output_feedback}

Both the transmission policies presented in previous sections
depend on the information from the state of the sensors 
and the controller-plant cascade. In this section
we relax this formulation, showing that the state of the controller-plant 
cascade can be replaced by an estimate, through a classical linear
continuous-time observer. We make the following assumption 
\begin{assumption} 
\label{assume:detectability}
 The pair $(A,C)$ in $\eqref{eq:plant}$ is detectable.
\end{assumption}

Considering the transmission-lazy closed-loop system in \eqref{eq:lazy_sys},
the introduction of an observer leads to the following formulation.
\begin{subequations}
\label{eq:olazy_sys}
\begin{align}
&	\left\{ \begin{array}{lll}
		\dot{x} &=& Ax + B \nu  \\
		\dot{\hat{x}} &=& A\hat{x} + B\nu + L(y-C\hat{x}) \\
		\dot{\nu} &=& 0 \\
		\dot{\tau} &=& \overline{1}-\dz(\frac{\tau}{\rho} ) \\
	 \end{array}\right. & 
	(\hat{x},\nu,\tau)\in \mathcal{C}_\Delta \\
&	\left\{\begin{array}{lll}
		x^+ &=& x  \\
		\hat{x}^+ &=& \hat{x} \\
		\nu^+ &=& g(\hat{x},\nu,\tau) \\
		\tau^+ &=& h(\hat{x},\nu,\tau) \\
	\end{array} \right. & 
        (\hat{x},\nu,\tau)\in \mathcal{D}_\Delta \\
&	 \begin{array}{lll}
		y &=& Cx.  
	 \end{array} 
\end{align}
\end{subequations}
where the flow dynamics is enriched by the observer dynamics
with gain $L\in\real^{n\times q}$,
and where $\hat{x}$ replaces $x$ within the functions $g$ and $h$.
Thus, looking at the definition of flow and jump sets in \eqref{eq:olazy_sys},
transmissions depend now on the state $\nu$ of the sensors and the 
estimate $\hat{x}$ of the controller-plant cascade state. 
The following stability results extend the result of Theorems 
\ref{thm:GAS_synchronous}, \ref{thm:GAS_asynchronous} 
to the output feedback case.

\begin{theorem}
\label{thm:GASoutput} 
 Under Assumption \ref{assume:detectability}, suppose that $(A+LC)$ is a Hurwitz matrix.
 \begin{enumerate}
  \item Consider the synchronous transmission policy of Section \ref{sec:synchronous}
  and suppose that the hypothesis of  Theorem \ref{thm:GAS_synchronous} are satisfied.
  Then there exists $0<\Delta<\rho$ (sufficiently small) such that the set
  \begin{equation}
  \label{eq:A_obs_sync} 
  \mathcal{A} := \{0\}\!\times\! \{0\} \!\times\! \{0\} \!\times\! [0,2\rho]
  \subset \realn\!\times\!\realn\!\times\!\real^q\!\times\!\real
  \end{equation}
  is globally asymptotically stable for the closed-loop system \eqref{eq:olazy_sys}.
  Moreover, if $B$ in \eqref{eq:plant} is full column rank, then $\mathcal{A}$ in \eqref{eq:A_obs_sync} is globally exponentially stable.
  \item Consider the asynchronous transmission policy of Section \ref{sec:asynchronous}
  and suppose that the hypothesis of  Theorem \ref{thm:GAS_asynchronous} are satisfied. 
  Then there exists $0<\Delta<\rho$ (sufficiently small) such that the set
  \begin{equation}
  \label{eq:A_obs_async} 
  \mathcal{A} := \{0\}\!\times\! \{0\} \!\times\! \{0\} \!\times\! [0,2\rho]^q
  \subset \realn\!\times\!\realn\!\times\!\real^q\!\times\!\real^q
  \end{equation}
  is globally asymptotically stable for the closed-loop system \eqref{eq:olazy_sys}.
  Moreover, if $B$ in \eqref{eq:plant} is full column rank, then $\mathcal{A}$ in \eqref{eq:A_obs_async} is globally exponentially stable.
 \end{enumerate}
\end{theorem}

{
The key point of Theorem \ref{thm:GASoutput} is in showing that the transmission policies do
not need any modification if we implement them by replacing the state of the
plant/controller cascade by an estimate. The proof of this fact is greatly simplified by
the adoption of the hybrid framework of \cite{GoebelCSM09},\cite{Goebel06}.}

 From the definition of $g$ in Sections
 \ref{sec:synchronous} and \ref{sec:asynchronous} and 
 looking at the jump dynamics of \eqref{eq:olazy_sys},
 each transmission is now based on estimate $\hat{y}:=C\hat{x}$ which replaces 
 the measured output $y=Cx$ and
 enforces a decoupled structure of the t-lazy closed-loop system. For instance, the policies 
 are now based on $\hat{x}$, through a comparison between the quantities $\nu-C\hat{x}$ and $\hat{x}$. 
 Thus, for example, the jump dynamics of the synchronous policy is now 
 given by $\nu^+= C\hat{x}$ which allows 
 for a mismatch dynamics at jumps given by 
 $\nu^+-C\hat{x}^+ = C\hat{x}-C\hat{x} =0$, paralleling the jumps dynamics of the 
 state-feedback case in which $\nu^+= y = Cx$ guarantees that $\nu^+ - Cx^+ = y-Cx=0$. Following this
 approach, the transmission policies operate on the subsystem $(\hat{x},\nu)$, whose state is available, 
 and the stability of the whole closed-loop system follows from the convergence of $\hat{x}$ to $x$,
 which is guaranteed by Assumption \ref{assume:detectability} and by the average dwell-time 
 between jumps enforced by the timers dynamics.

\begin{proofof}{\emph{Theorem \ref{thm:GASoutput}}.}
{
We extend the argument of the proofs of Theorems \ref{thm:GAS_synchronous} 
and \ref{thm:GAS_asynchronous} to the observer dynamics. The proof is 
divided in two parts: the first one concerns the analysis of the synchronous case,
while the second one develops the analysis of the asynchronous case. 
For each case, we first
consider a generalized error system. Then, we give a Lyapunov function and we
show that along the solutions to the hybrid system it satisfies several 
inequalities on the jump and flow dynamics. These
inequalities are used in combination to an invariance principle to show
global asymptotic stability. Finally, we strengthen these results to 
exponential bounds by relying on the presence of a dwell-time.}

{ \underline{Synch, error dynamics and Lyapunov function:}} 
using the coordinate transformation 
 $(\hat{x},e,\eta) = (\hat{x},\nu-C\hat{x},x-\hat{x})$ and 
 considering the \emph{synchronous transmission policy} of Section \ref{sec:synchronous},
 we can rewrite \eqref{eq:olazy_sys} as follows 
\begin{equation}
\label{eq:olazy_sys_synch}
  \begin{array}{ll} 
	\left\{ \begin{array}{lll}
		\dot{\hat{x}} &=& F_{11}\hat{x} + F_{12}e + LC\eta\\
		\dot{e} &=& F_{21}\hat{x} + F_{22}e -CLC\eta \\
		\dot{\eta} &=& (A-LC)\eta \\
		\dot{\tau} &=& \overline{1}-\dz(\frac{\tau}{\rho} ) \\
	 \end{array}\right. & 
	\;(\hat{x},e)\!\in\! \overline{\mathcal{C}} \mbox{ or }   0 \!\leq\! \tau \!\leq\! \Delta \\
	\left\{\begin{array}{lll}
		\hat{x}^+ &=& \hat{x} \\
		e^+ &=& 0 \\
		\dot{\eta}^+ &=& \eta\\
		\tau^+ &=& 0 \\
	\end{array} \right. & 
	\;(\hat{x},e)\!\in\! \overline{\mathcal{D}} \mbox{ and }  \tau \!\geq\! \Delta \\
 \end{array}
\end{equation}
Using the aggregate state $X:= \smallmat{\hat{x}^T, e^T, \tau}^T$ and 
$Y:= \smallmat{\hat{x}^T, e^T, \tau, \eta}^T$, 
consider $W(X)$ defined in \eqref{eq:W(X)},
and define $P_o=P_o^T>0$ such that $(A+LC)^TP_o + P_o(A+LC) \leq -I$, from which we can define the 
Lyapunov function $V:\realn\times\real^q\times\real\times \realn \to \real_{\geq 0}$ given by 
\begin{equation}
\label{eq:proofthm3_0}
 V(Y) := W(X) + \gamma \eta^T P_o \eta. 
\end{equation}
where $\gamma>0$.
From the definition of $V$, using the bounds on $W$, 
there exists $\underline{\alpha},\overline{\alpha}>0$ such that 
$\underline{\alpha}|Y|^2_\mathcal{A} \leq V(Y) \leq \overline{\alpha}|Y|^2_\mathcal{A}$. 

{ \underline{Synch, Lyapunov function at jumps:}} 
using \eqref{eq:proofthm1_a}, we have 
\begin{equation}
\label{eq:proofthm3_a}
 V(Y^+) - V(Y)\leq -e^TP_2e  
\end{equation}
for each $Y$ such that $(\hat{x},e)\in\overline{\mathcal{D}}$ and $\tau\geq \Delta$. 

{ \underline{Synch, Lyapunov function on flows:}} 
(i)~for $0 \leq \tau \leq \Delta$, using \eqref{eq:proofthm1_b} and $\lambda_0$, $\lambda_1$
given in the last inequality of \eqref{eq:proofthm1_b}, pick $\lambda$ and $\rho$ as in 
the proof of Theorem \ref{thm:GAS_synchronous} and define
$\gamma_0 := \frac{\lambda_0}{2}$, 
$\gamma_1 := \lambda_1\exp((2\rho-\Delta)\lambda)$, 
$\gamma_2 := 2|P_1 L C|$, 
$\gamma_3 := 2\exp(2\rho\lambda)|P_2 C L C |$.
Then, we get 
\begin{equation}
\label{eq:proofthm3_b}
 \begin{array}{lll}
  \dot{V}(Y) 
&\leq&  -\gamma_0 |\hat{x}|^2 - \gamma_1|e|^2 + 2\hat{x}^T P_1LC\eta \\
& &- 2\exp((2\rho-\tau)\lambda)e^TP_2CLC\eta -\gamma|\eta|^2 \\
&\leq&  -\gamma_0 |\hat{x}|^2 - \gamma_1|e|^2 + \gamma_2|\hat{x}||\eta| + \gamma_3|e||\eta| -\gamma|\eta|^2 \\
&\leq&  -\frac{\gamma_0 }{2} |\hat{x}|^2 -\frac{\gamma_1 }{2} |\hat{e}|^2 -\frac{\gamma }{2} |\eta|^2 \vspace{-3mm}
 \end{array} \vspace{1mm}
\end{equation}
where the last inequality is established by using 
$\varepsilon  a^2 + \frac{1}{\varepsilon}b^2 \geq ab \ a,b \in \real_{\geq 0},  \varepsilon>0$, 
for $\gamma>0$ sufficiently large.

(ii)~For $\tau \geq \Delta$, $(\hat{x},e) \in \overline{\mathcal{C}}$, 
using \eqref{eq:proofthm1_c} with $\rho$ sufficiently small, we get
\begin{equation}
\label{eq:proofthm3_c}
 \begin{array}{lll}
  \dot{V}(Y) 
&\leq&  -\frac{\gamma_x }{2} |\hat{x}|^2 + \gamma_2|\hat{x}||\eta| + \gamma_3|e||\eta| -\gamma|\eta|^2 \\
&\leq&  -\frac{\gamma_x }{2} |\hat{x}|^2 + (\gamma_2 + \gamma_3\gamma_e)|\hat{x}||\eta| -\gamma|\eta|^2 \\
&\leq&  -\frac{\gamma_x }{4} |\hat{x}|^2 -\frac{\gamma }{2} |\eta|^2 
 \end{array}
\end{equation}
where, as before, the last inequality holds for $\gamma>0$ sufficiently large.

{ \underline{Synch, GAS of $\mathcal{A}$ by invariance principle:}} 
from the inequalities above we can establish global asymptotic stability of the set $\mathcal{A}$
following the argument of the proof of Theorem \ref{thm:GAS_synchronous}.
For instance, define 
$\ell(\mu) := \{Z\,|\,\,W(Z)=\mu\}$. 
For each solution $Y:=(\hat{x},\eta,e,\tau)$ such that $Y(0,0) \in \ell_\mu$, we have that  
(i)~on flows, when  $\hat{x}, \eta \neq 0$, $V$ decreases; 
(ii)~ on flows, when $\hat{x} = 0$ and $\eta =0$, either $V$ decreases 
or $(\hat{x},e)\in \overline{\mathcal{C}}$, from which $e=0$, that is, $V(Y) =0$; 
(iii)~on jumps, $V$ does not increase but after each jump the system must flow for a $\Delta$ interval 
of time thus, necessarily, $\ell(\mu)$ is not an invariant set. 
global asymptotic stability follows from \cite[Theorem 23]{GoebelCSM09}.

{ \underline{Synch, exponential stability of the set $\mathcal{A}$:}}
exponential stability of the set $\mathcal{A}$ in \eqref{eq:A_obs_sync}
can be established by using \cite[Theorem 2]{TeelTAC13},
as in the proof of Theorem \ref{thm:GAS_synchronous}. In fact, 
the pair 
\begin{equation}
\label{eq:pair_obs}
\Big(\smallmat{
I_n & 0 & 0 \\
0 & 0 & I_n
},
\smallmat{
F_{11} & F_{12} & LC \\
F_{21} & F_{22} & CLC \\
0 & 0 & A-LC
}
\Big) 
\end{equation}
is observable when $B$ is full column rank 
(by linear transformation and PBH-test).
Thus, decomposing the state  
in $\xi_1 = (\hat{x},e,\eta)$ and $\xi_2 = \tau$,
using \eqref{eq:proofthm3_a}-\eqref{eq:proofthm3_c},
and observing that 
$(t,j)\in \dom\,X$ implies $j \leq \frac{t}{\Delta}$,
every condition of \cite[Assumption 1]{TeelTAC13} is satisfied.
Therefore, $\mathcal{A}$ is globally exponentially stable
from \cite[Theorem 2]{TeelTAC13}.

{ \underline{Asynch, error dynamics:}} 
for the \emph{asynchronous transmission policy}, \eqref{eq:olazy_sys} becomes 
\begin{equation}
\label{eq:olazy_sys_asynch}
  \begin{array}{ll} 
	\left\{ \begin{array}{lll}
		\dot{\hat{x}} &=& F_{11}\hat{x} + F_{12}e + LC\eta\\
		\dot{e} &=& F_{21}\hat{x} + F_{22}e -CLC\eta \\
		\dot{\eta} &=& (A+LC)\eta \\
		\dot{\tau} &=& \overline{1}-\dz(\frac{\tau}{\rho} ) \\
	 \end{array}\right. & 
	 \begin{array}{l}
	  \quad\forall i\,\big(
	  (\hat{x},e_i)\!\in\! \overline{\mathcal{C}}_i \vspace{-1mm}\\
	  \quad\qquad\mbox{ or } 0 \!\leq\! \tau_i \!\leq\! \Delta
	  \big) \\	  
	 \end{array} \\
	\left\{\begin{array}{lll}
		\hat{x}^+ &=& \hat{x} \\
		e^+ &=& g(\hat{x},e+C\hat{x},\tau) -C\hat{x}\\ 
		\dot{\eta}^+ &=& \eta\\
		\tau^+ &=& h(\hat{x},e+C\hat{x},\tau)\\
	\end{array} \right. & 
	 \begin{array}{l}
	  \quad\exists i\,\big(
	  (\hat{x},e_i)\!\in\! \overline{\mathcal{D}}_i \vspace{-1mm}\\
	  \quad\qquad\mbox{ and } \tau_i \!\geq\! \Delta
	  \big) \\	  
	 \end{array} \vspace{-2mm}\\ 
 \end{array} \vspace{1mm}
\end{equation}
where $g$ and $h$ are defined in \eqref{eq:g_async} and \eqref{eq:h_async}.

{ \underline{Asynch, Lyapunov inequalities:}} 
using $V$ in \eqref{eq:proofthm3_0}. at jumps we get
\begin{equation}
\label{eq:proofthm3_d}
 V(Y^+) - V(Y)\leq -p_i |e_i|^2  
\end{equation}
while on flows, using \eqref{eq:proofthm2_f} and \eqref{eq:proofthm2_d}
with an argument similar to the sequence of inequalities 
\eqref{eq:proofthm3_b} and \eqref{eq:proofthm3_c},
we get the following inequality
\begin{equation}
\label{eq:proofthm3_e}
 \begin{array}{lll}
  \dot{V}(Y) 
&\leq&  -\gamma_1 |\hat{x}|^2 -\gamma_2 |\eta|^2  
        -\gamma_3 \sum\nolimits\limits_{i\in \mathcal{I}_\Delta} |e_i|^2
 \end{array}
\end{equation}
where $\mathcal{I}_{\Delta}:=\{i\,|\, \tau_i \leq \Delta\}\subseteq \{1,\dots,q\}$,
for some $\gamma_1,\gamma_2,\gamma_3 >0$.

{ \underline{Asynch, global asymptotic and exponential stability:}} 
using inequalities \eqref{eq:proofthm3_d} and \eqref{eq:proofthm3_e},
the definition of the sets $\overline{\mathcal{C}}_i$ and $\overline{\mathcal{D}}_i$,
and the fact that for any given solution $Y$, if $(t,j)\in\dom\,Y$ then 
$j\leq \frac{qt}{\Delta}$, we can establish that the set $\ell(\mu)$ is not invariant
for any given $\mu>0$, from which global asymptotic stability of the set $\mathcal{A}$
in \eqref{eq:A_obs_async} follows by \cite[Theorem 23]{GoebelCSM09}.
Finally, exponential stability follows from \cite[Theorem 2]{TeelTAC13}.
In fact, conditions 1)-4) of \cite[Assumption 1]{TeelTAC13} can be established 
as shown above for the synchronous policy, while condition 5) of \cite[Assumption 1]{TeelTAC13}
follows from the fact that $(t,j)\in\dom\,Y$ implies $j\leq \frac{qt}{\Delta}$, for any given solution $Y$. 
\end{proofof}

{

\section{Robustness of the transmission policies}
\label{sec:robustness}

A fundamental feature of the proposed hybrid model \eqref{eq:lazy_sys} or
\eqref{eq:olazy_sys} is that asymptotic stability is robust. 
In fact, these two models satisfy the
so-called \emph{basic conditions} \cite{GoebelCSM09}, recalled in Section~\ref{sec:preliminaries}, which guarantee
several interesting regularity properties of the space of solutions 
to the hybrid system. This regularity is exploited to establish 
several robustness results for hybrid systems stability.
Based on these results, we show in this section that 
the stability proven in Theorems \ref{thm:GAS_synchronous},
\ref{thm:GAS_asynchronous}, and \ref{thm:GASoutput} is indeed robust.
This is one of the main advantages of modeling the sampling transmission
policies within the hybrid systems framework proposed of 
\cite{Goebel06},\cite{GoebelCSM09}.

We will not enter into the details of robustness theory for hybrid systems.
The interested reader is referred to \cite{Goebel12}. 
Instead, we will show how to characterize parameter perturbations, 
measurements noise, and transmission delays
as a perturbed hybrid model $\mathcal{H}_\delta$ where $\delta$ represents a 
perturbation radius with respect to the original system $\mathcal{H}$. 
Then, for example, we may invoke \cite[Theorem 17]{GoebelCSM09} 
to establish that the asymptotic stability enforced by 
the transmission policies without perturbation turns to 
practical stability when parameters perturbations, measurement noise,
and delays are sufficiently small. 
We do not propose any formal statement here. 
We will keep the exposition at level of a discussion.

Robustness to \emph{measurement noise}, \emph{parameter uncertainties} and \emph{transmission delays} 
can be characterized as shown in \cite[p.57]{GoebelCSM09}. For simplicity of exposition, let 
us consider only the error model of the synchronous case \eqref{eq:elazy_sys}. The analysis of the
other cases \eqref{eq:elazy_sys2}, \eqref{eq:olazy_sys_synch}, \eqref{eq:olazy_sys_asynch} 
is very similar. As a first step, let us rewrite \eqref{eq:elazy_sys} 
in the following simplified form
\begin{equation}
\label{eq:rob_sys}
  \begin{array}{ll} 
	\left\{ \begin{array}{lll}
		\dot{x} &=& f_1(x,e)\\
		\dot{e} &=& f_2(x,e) \\
		\dot{\tau} &=& f_3(\tau) \\
	 \end{array}\right. & 
	 \begin{array}{l}
	  \quad
	  (x,e)\!\in\! \overline{\mathcal{C}} \vspace{-1mm}\\
	  \quad\qquad\mbox{ or } 0 \!\leq\! \tau \!\leq\! \Delta \\	  
	 \end{array} \\
	\left\{\begin{array}{lll}
		x^+ &=& x \\
		e^+ &=& 0\\ 
		\tau^+ &=& 0\\
	\end{array} \right. & 
	 \begin{array}{l}
	  \quad
	  (x,e)\!\in\! \overline{\mathcal{D}} \vspace{-1mm}\\
	  \quad\qquad\mbox{ and } \tau \!\geq\! \Delta \\	  
	 \end{array} \vspace{-2mm}\\ 
 \end{array} \vspace{1mm}
\end{equation}
where the definition of $f_1$, $f_2$ and $f_3$ is clear from a comparison to \eqref{eq:elazy_sys}.
Then, we can address different kinds of perturbations by considering the following 
perturbed hybrid model.
\begin{equation}
\label{eq:rob_sys1}
  \begin{array}{ll} 
	\left\{ \begin{array}{lll}
		\dot{x} &\in& f_1(x,e) + \delta_1 \mathbb{B}\\
		\dot{e} &\in& f_2(x,e) + \delta_2 \mathbb{B}\\
		\dot{\tau} &\in& f_3(\tau+d_1) + \delta_3 \mathbb{B}\\
	 \end{array}\right. & 
	 \begin{array}{l}
	  \quad
	  (x+d_2,e+d_3)\!\in\! \overline{\mathcal{C}}+ \delta_4 \mathbb{B} \vspace{-1mm}\\
	  \quad\qquad\mbox{ or } 0 \!\leq\! \tau \!\leq\! \Delta \\	  
	 \end{array} \\
	\left\{\begin{array}{lll}
		x^+ &=& x \\
		e^+ &=& 0+d_4\\ 
		\tau^+ &=& 0\\
	\end{array} \right. & 
	 \begin{array}{l}
	  \quad
	  (x+d_5,e+d_6)\!\in\! \overline{\mathcal{D}}. \vspace{-1mm}\\
	  \quad\qquad\mbox{ and } \tau \!\geq\! \Delta \\	  
	 \end{array} \vspace{-2mm}\\ 
 \end{array} \vspace{1mm}
\end{equation}
where $d_i$, $i\in\{1,\dots,6\}$ are disturbance signals, and
$\delta_i>0$ are constants. Note that $f_i(x,e) + \delta_i \mathbb{B}$ 
is now a set-valued map for each $i\in\{1,2,3\}$.

We can identify the following three cases of interest.
\begin{itemize}
\item Consider $\delta_1=\delta_2=\delta_3=\delta_4 = 0$ and suppose that
The signal $d_1,\dots, d_6$ are typically small, thus 
bounded within a ball of radius $\delta_0 \mathbb{B}$, $\delta_0 >0$. 
They characterize 
possible \emph{measurement noise} that perturbs the system. Respectively,
$d_1$ enforces a drift on the internal timer of the t-lazy sensors, 
$d_2,d_3,d_5,d_6$ introduce a perturbation on the transmission decision, 
$e=0+d_4$, that is, $v=y+d_4$ corrupts the transmitted
sample.

\item Suppose $d_i=0$ for each $i\in\{1,\dots,6\}$, $\delta_4=0$,
and $\delta_i>0$ for each $i\in\{1,2,3\}$. 
$f_i(x,e) + \delta_i \mathbb{B}$ is a (outer semicontinuous, convex and not-empty)
set valued mapping of radius $\delta_i$, centered at $f_i(x,e)$. 
The use of set-valued mappings is a possible approach to the characterization
of \emph{parameters uncertainty} on the model.

\item Suppose $d_i=0$ for each $i\in\{1,\dots,6\}$, 
$\delta_i=0$ for each $i\in\{1,2,3\}$, and $\delta_4>0$.
Since the flow set is now larger, also the space of solutions
of the perturbed system is larger than the
space of solutions of the original hybrid model \eqref{eq:elazy_sys}.
The new solutions have an "excess" of flow which exactly characterizes
\emph{transmission delays}. In fact, the excess of flow represents a 
scenario in which a sample transmission should be performed but the
system continues to flow for a given (small) amount of time.
\end{itemize}

Using the fact that the original model \eqref{eq:elazy_sys} satisfies
the basic conditions (see Section~\ref{sec:preliminaries}), the continuity of the flow map in \eqref{eq:elazy_sys},
and the fact that each signal $d_i$ is bounded by $\delta_0\mathbb{B}$,
we invoke \cite[Theorem 17]{GoebelCSM09} (as done in \cite[p.57]{GoebelCSM09})
to establish that for any given compact set of initial conditions $\mathcal{K}$
and any $\varepsilon>0$,
there exist sufficiently small values of $\delta_i>0$, $i\in\{0,1,2,3,4\}$,
such that all the solutions to \eqref{eq:rob_sys}
asymptotically (and uniformly) converge to the set 
$\mathcal{A}+ \varepsilon \mathbb{B}$.

It follows that the synchronous transmission policy is robust to (small) measurement noise,
parameters uncertainty and transmission delays. 
This property corresponds to semiglobal practical stability, where
\emph{practical} refers to the fact that solutions converge to a neighborhood
of $\mathcal{A}$ given by $\mathcal{A}+ \varepsilon \mathbb{B}$, while \emph{semiglobal} refers
to the fact that the initial compact set $\mathcal{K}$ can be taken arbitrarily large.
Note that similar arguments can be used to establish robustness of each policy presented
in this paper.
}

{
\begin{remark}
While this section illustrates the robustness of our scheme for small
disturbances and/or perturbations, it is also of interest to
characterize some level of robustness in the presence of large seldom
events, such as packet corruption or loss within the transmission
network. This phenomenon can be captured by adding an extra state to
(\ref{eq:lazy_sys_flow}), (\ref{eq:lazy_sys_jump}), which acts like an
 input matched disturbance to
the right hand side of the first equation in (\ref{eq:lazy_sys_flow})
and remains constant between jumps. Robustness to this (non-small)
disturbance can then be addressed by making assumptions on its
persistency (for example one may use dwell-time or average dwell-time
assumptions between packet losses, that is nonzero occurrences of this
disturbance) and exploiting the strict decrease given by our Lyapunov
functions to dominate the effect of this large but not persistent
disturbance.
Developing this analysis is beyond the scope of this paper.
\end{remark}
}

\section{Simulation examples} 
\label{sec:example} 

\subsection{State-feedback and output feedback}
\label{sec:example1} 

Consider an unstable plant given by the transfer function $ \frac{s+2}{(s+1)(s-3)} $,
which can be stabilized by negative static output feedback, for example by using the gain $k = 9$.
The controller-plant cascade can be represented by
the following state space equations
\begin{equation}
 \label{exampleA:plant}
\left\{
\begin{array}{lll}
 \dot{x} & = & \smallmat{ 2 & 1.5 \\ 2 & 0 } x + \smallmat{ -18 \\ 0 } u_c \\
  y_p & = & \smallmat{ 0.5 & 0.5}x,
\end{array} \right.
\end{equation}
and the nominal closed-loop system is given by \eqref{exampleA:plant} 
through the interconnection $u_c = y_p$.

\begin{figure}[ht!]
\begin{center}
\includegraphics[width=0.49\columnwidth]{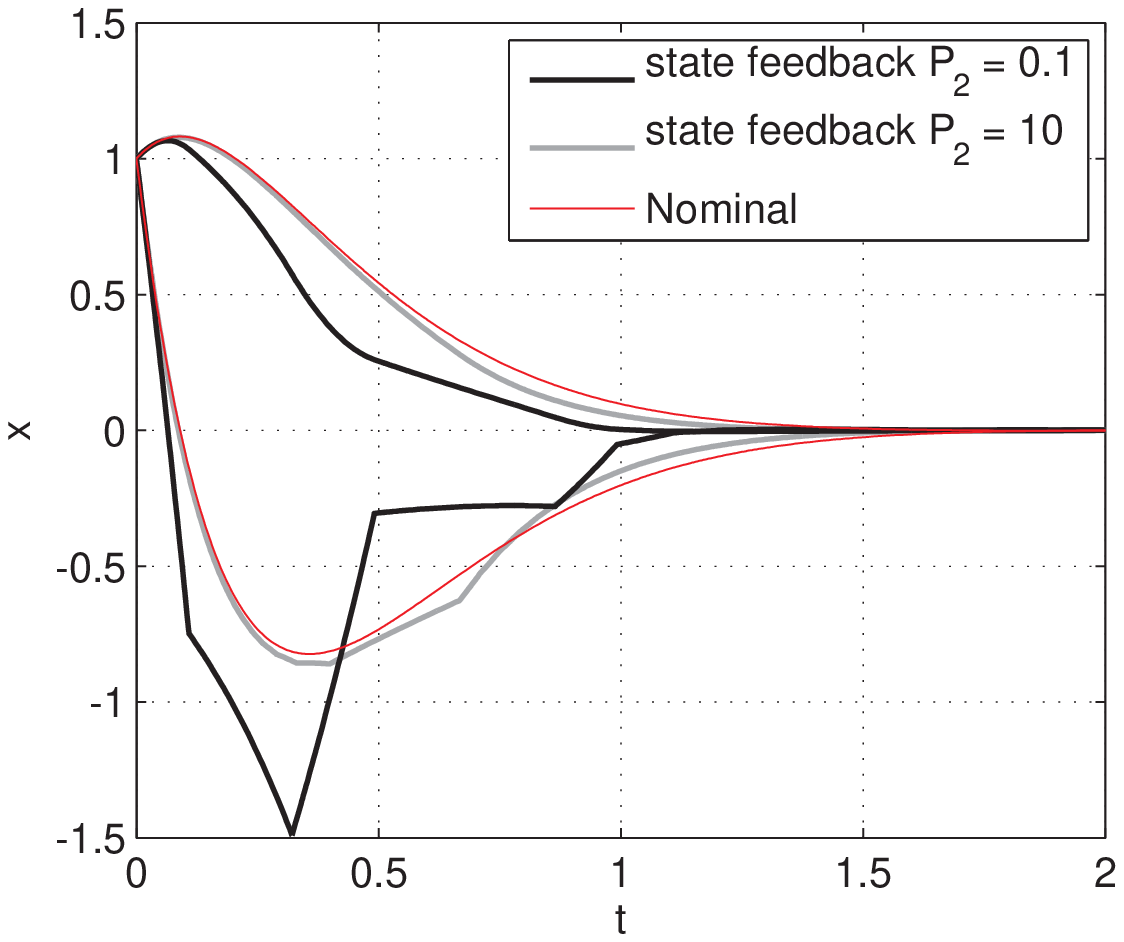}
\includegraphics[width=0.49\columnwidth]{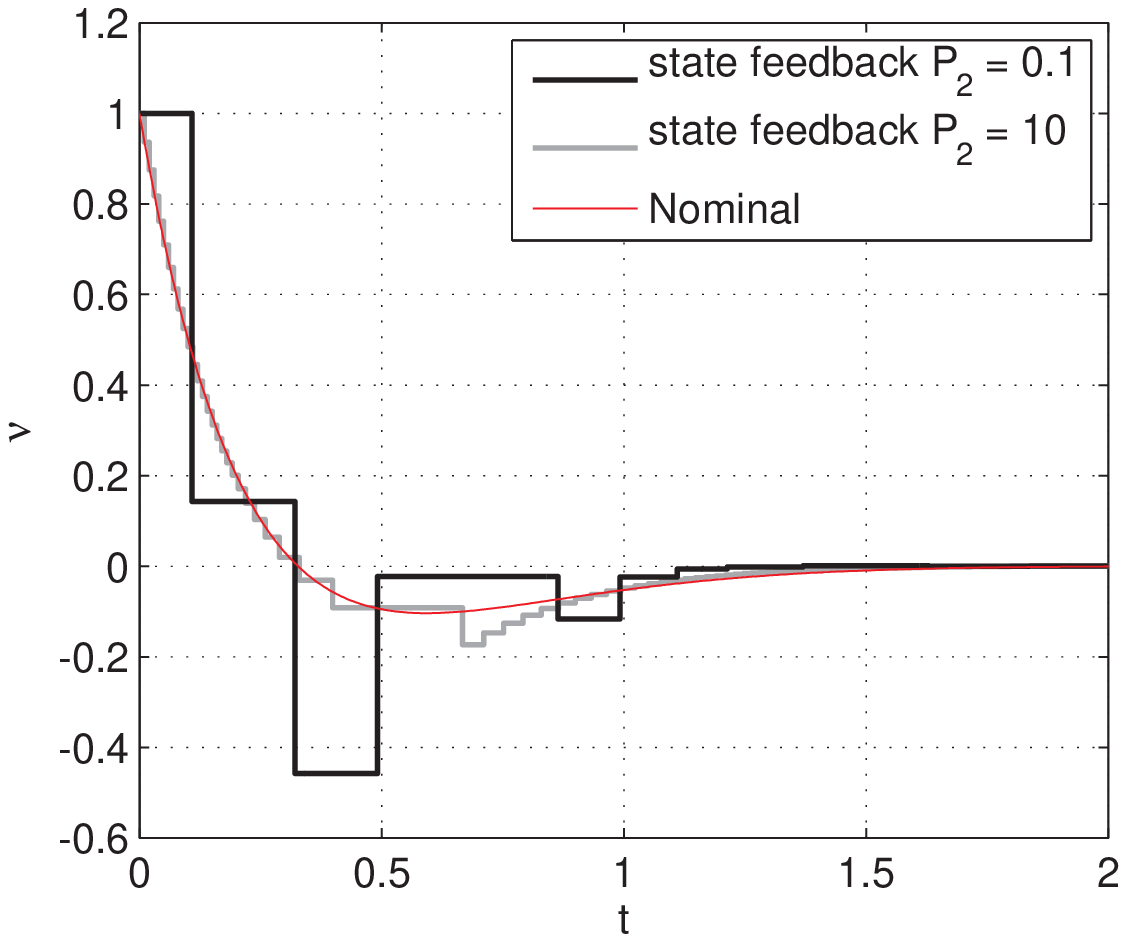}
\includegraphics[width=0.49\columnwidth]{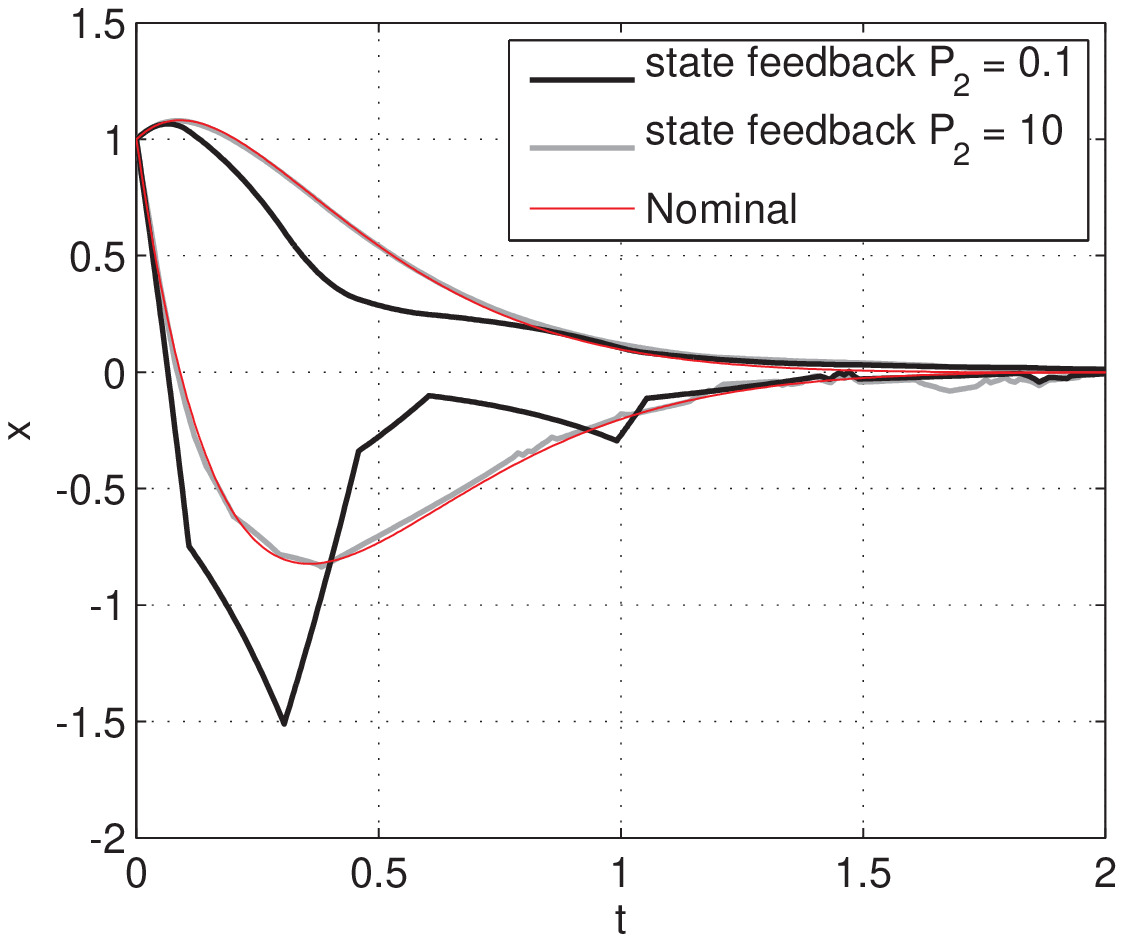}
\includegraphics[width=0.49\columnwidth]{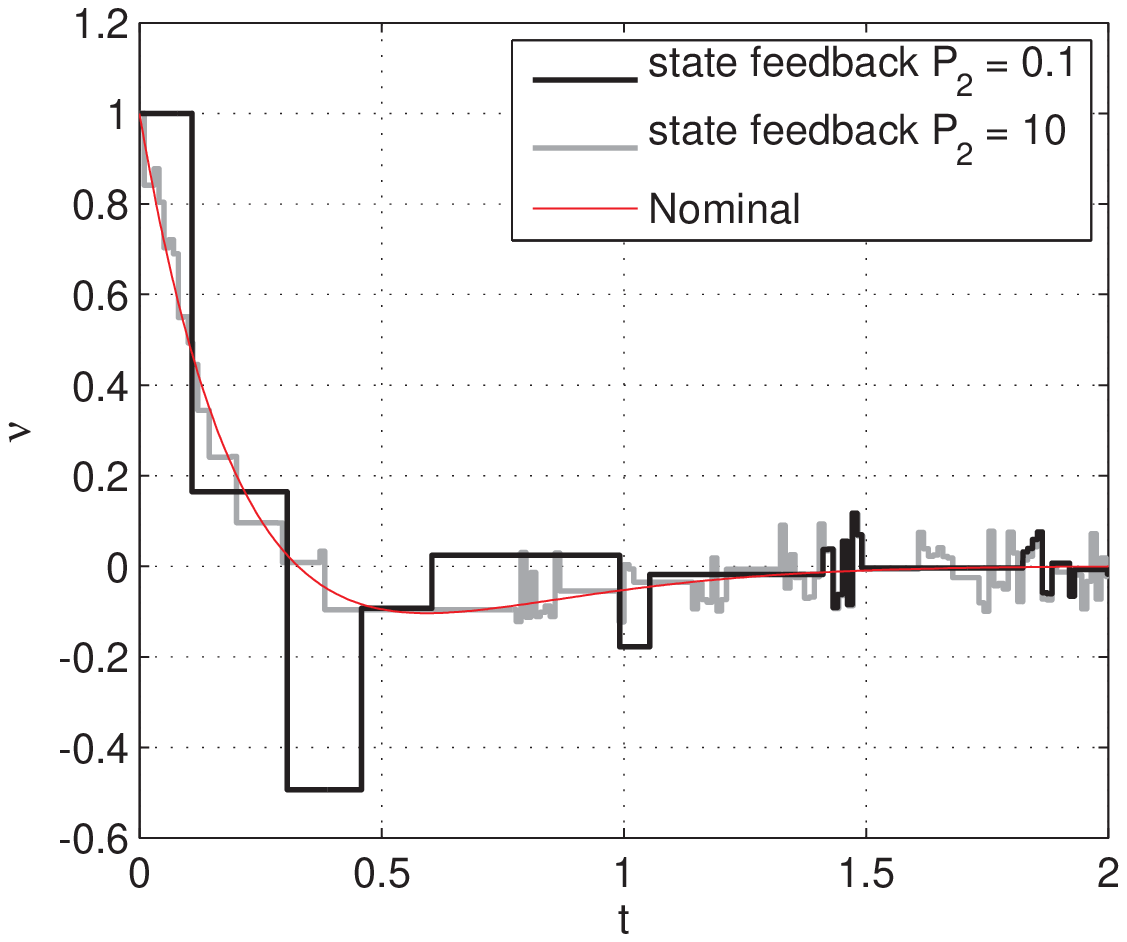}
\caption{Synchronous policy from state feedback: time evolution of
state $x$ and input $u_c$ for $P_2\in\{0.1,10\}$.
TOP - no sample corruption. BOTTOM - sample corruption
by uniform distributed noise between $-0.1$ and $0.1$.}
\label{exampleA:fig1}
\end{center}
\end{figure}

\begin{figure}[ht!]
\begin{center}
\includegraphics[width=0.49\columnwidth]{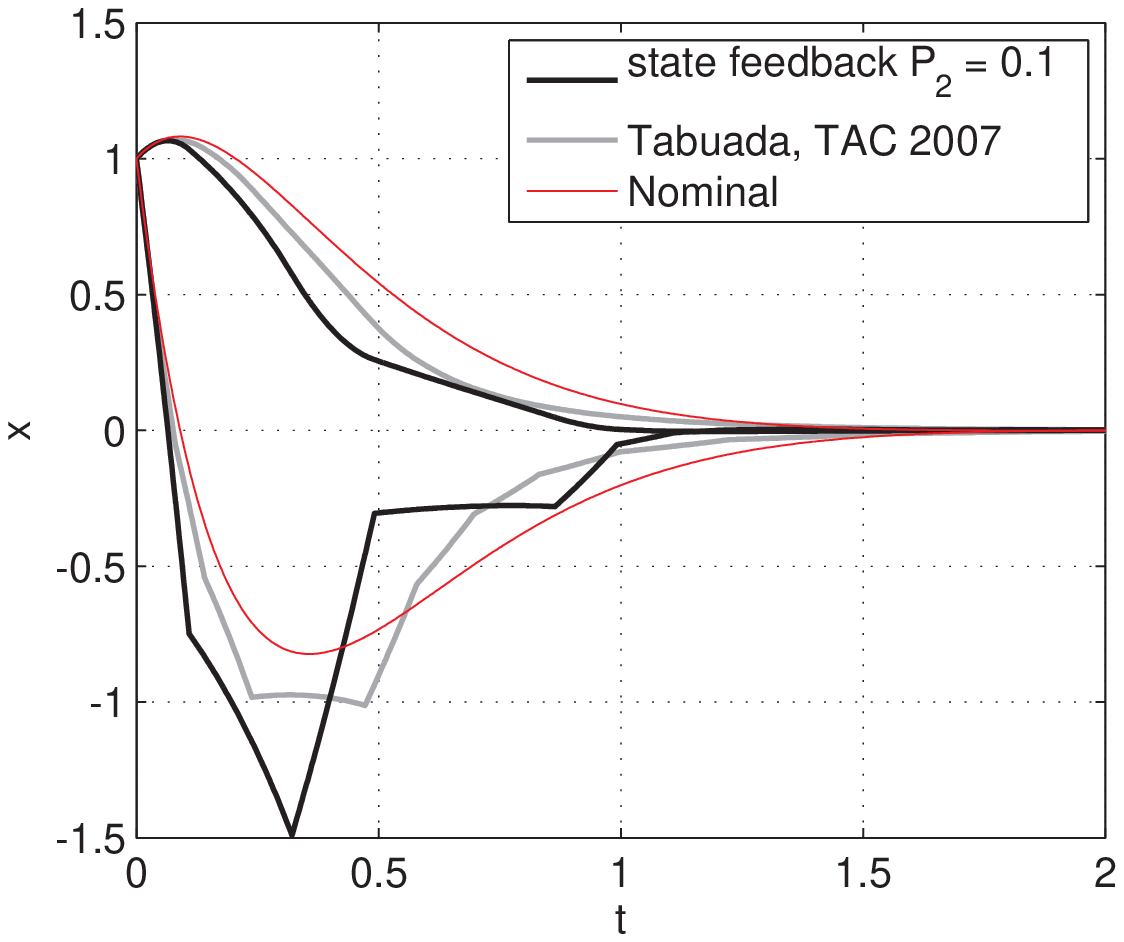}
\includegraphics[width=0.49\columnwidth]{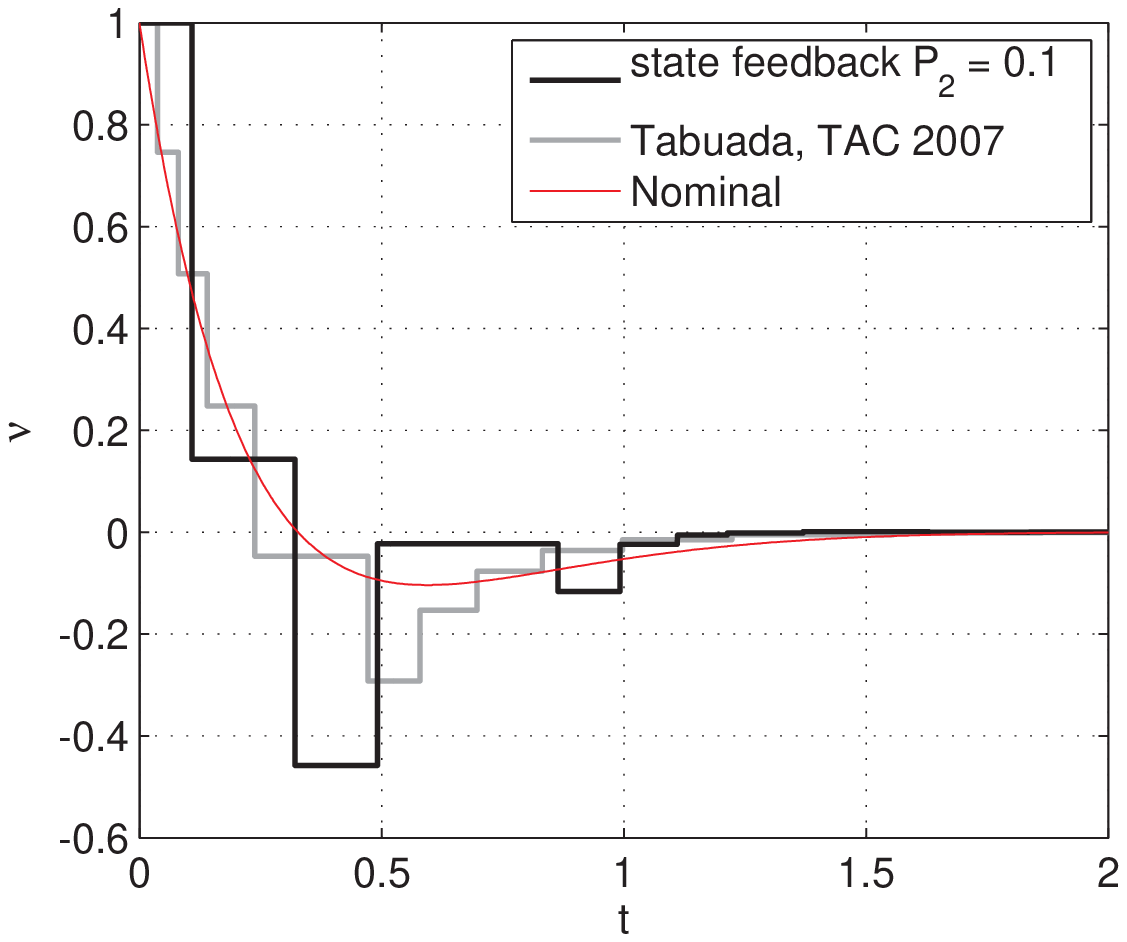}
\includegraphics[width=0.49\columnwidth]{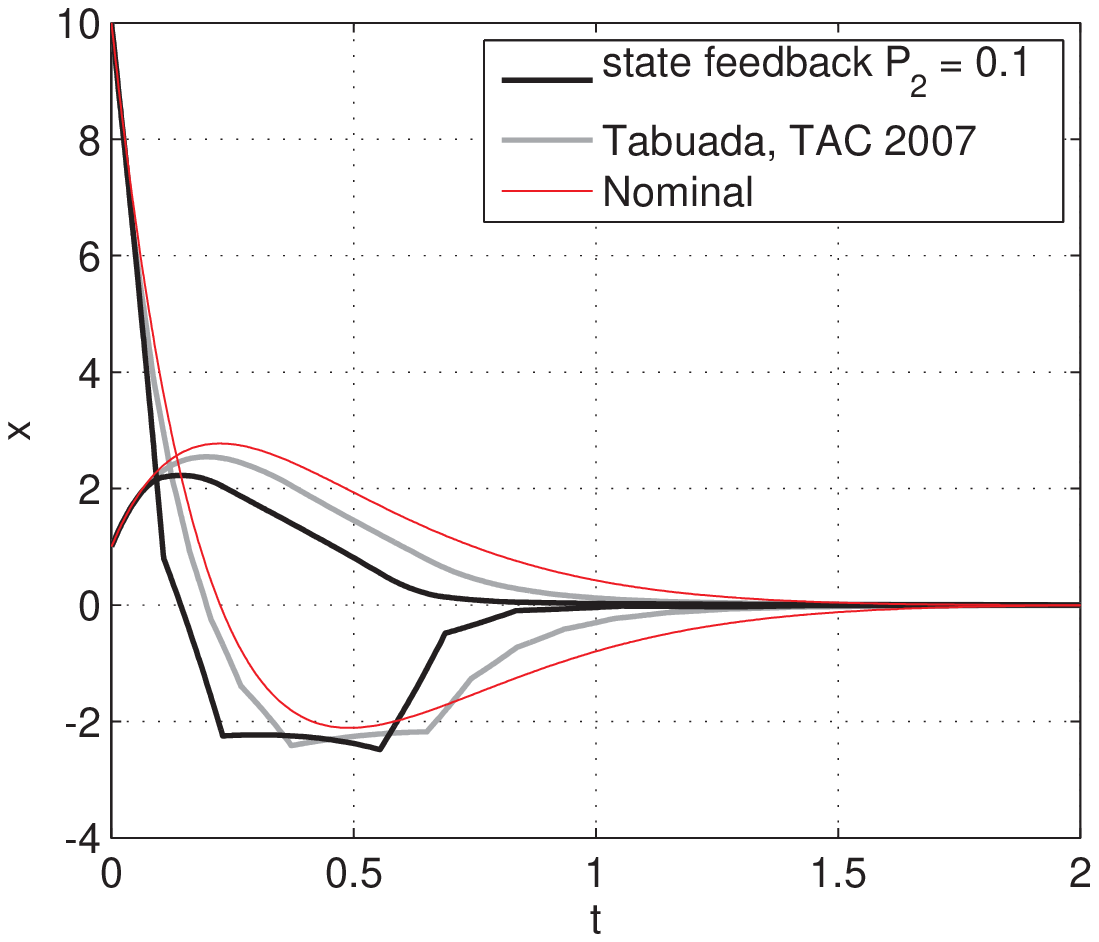}
\includegraphics[width=0.49\columnwidth]{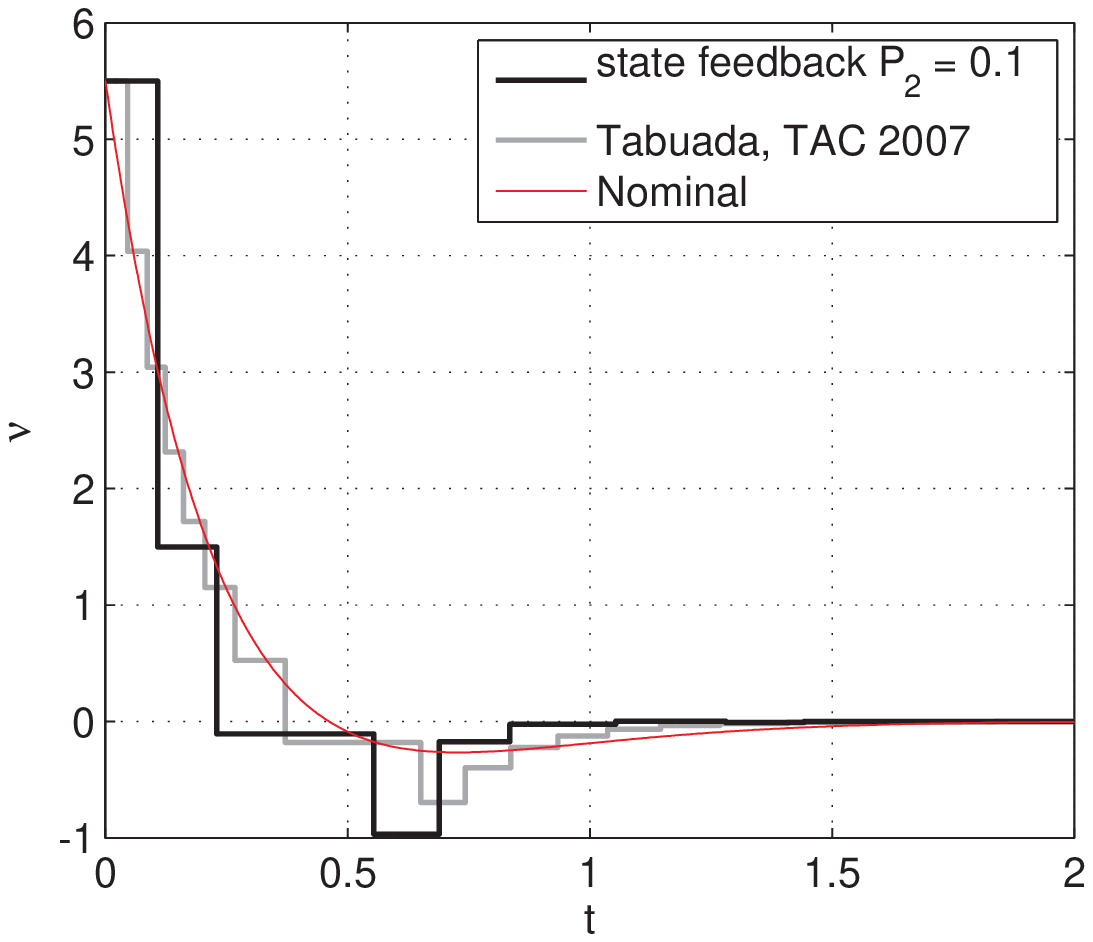}
\caption{Comparison between the state-feedback synchronous policy for
$P_2 = 0.1$
and the policy of \cite{Tabuada07}.
TOP - comparison from the (balanced) initial condition $x(0,0) = [1\,\,1]^T$.
BOTTOM - comparison from the (unbalanced) initial condition $x(0,0) = [10\,\,1]^T$.}
\label{exampleA:fig1bis}
\end{center}
\end{figure}

\begin{figure}[ht!]
\begin{center}
\includegraphics[width=0.9\columnwidth]{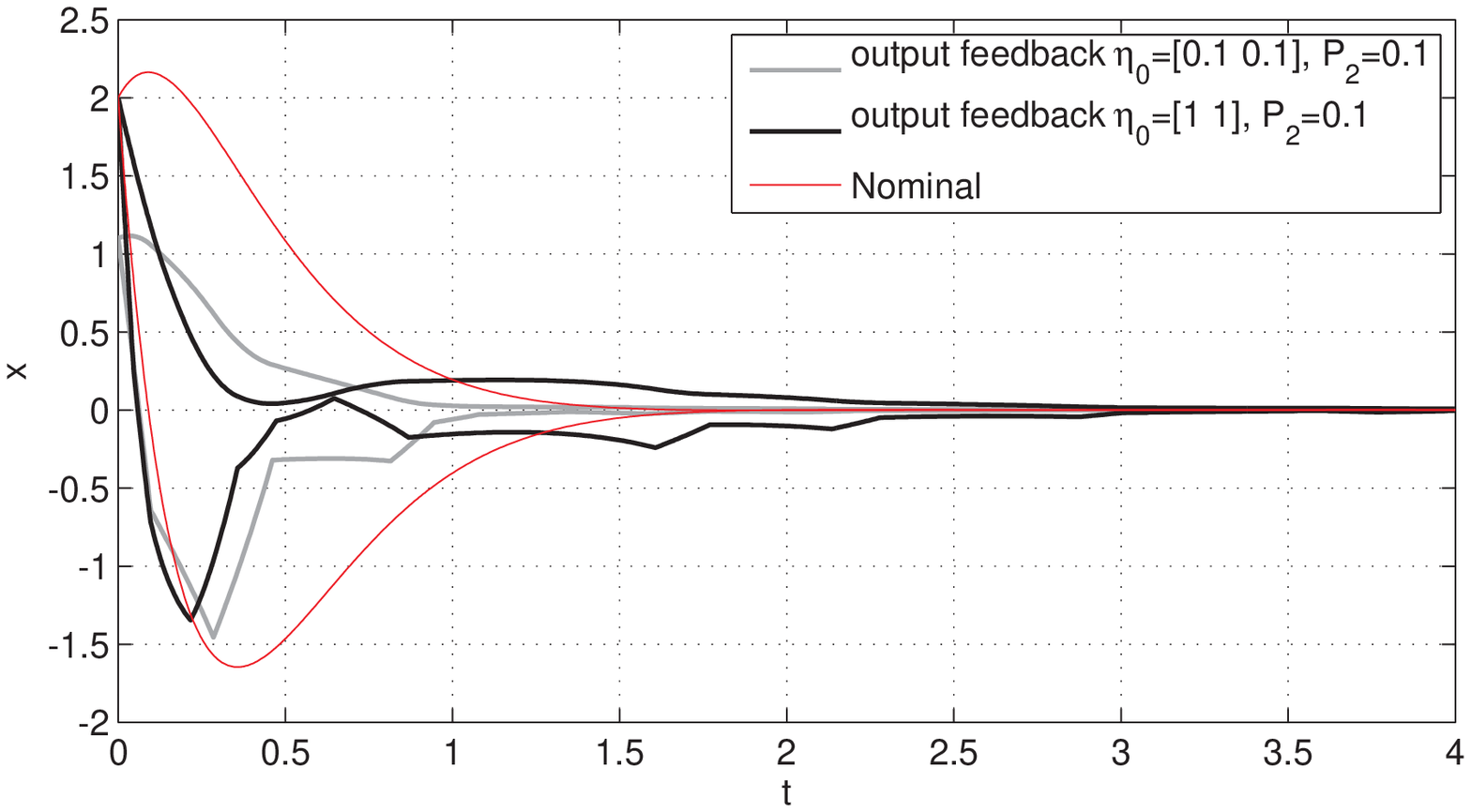}
\includegraphics[width=0.9\columnwidth]{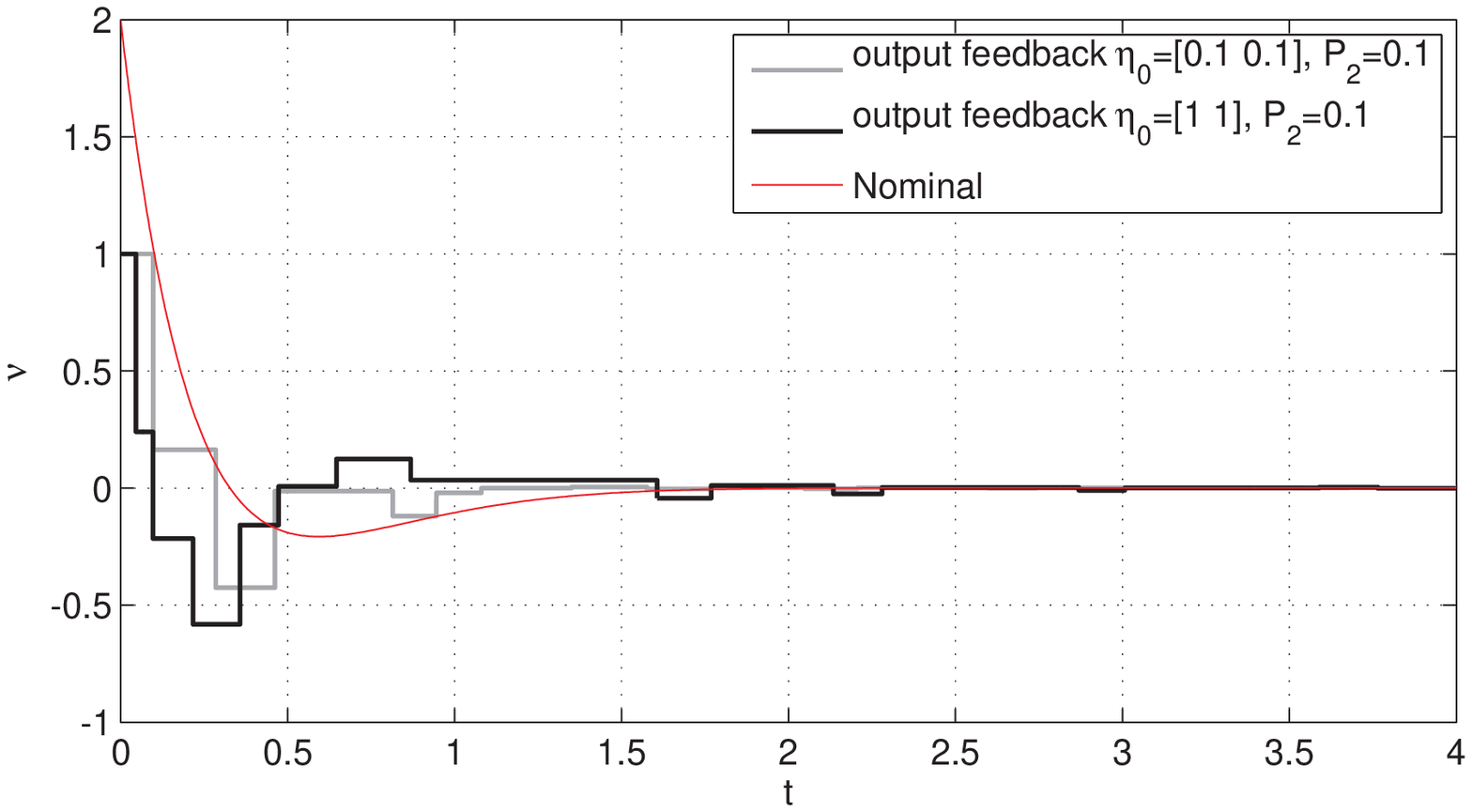}
\caption{Synchronous policy from output feedback: time evolution of
state $x$ and input $u_c$. $P_2=0.1$. Initial conditions:
$x(0,0) = [1\,\,1]^T$, and
$\eta(0,0) = [0.1\,\,0.1]^T$ - small estimation error (gray line), or 
$\eta(0,0) = [1\,\,1]^T$ - large estimation error (black line).  }
\label{exampleA:fig2}  
\end{center}
\end{figure}

{

We consider here the synchronous case. For
$P_1 = \smallmat{ 0.091 & 0.067 \\ 0.067 &   0.573}$, $0< \gamma_x \ll 1$,
$\gamma_e \gg 1$, and $P_2 \in \{0.1, 10\}$, conditions \textbf{(S1)} of 
Section \ref{sec:synchronous} are satisfied and the effect of the t-lazy sensors
on the trajectories of the system in the state-feedback case 
is summarized in Figures~\ref{exampleA:fig1}
and~\ref{exampleA:fig1bis}. 
In all the simulations the state $\nu$ is initialized at zero for simplicity.
The top row
of Figure \ref{exampleA:fig1} represents the time evolution of the
synchronous policy ($x$ and $\nu$) with state information available and
without corruption of the transmitted samples 
for $P_2 =0.1$ (black curve) and $P_2 =10$ (gray curve).
Note that the solid curve, corresponding to a smaller value of $P_2$
results in a significantly reduced data rate, as compared to the other
selection (gray) whose trace is almost coincident with the nominal
closed-loop (thin red curve), namely the closed-loop with no transmission
channel. 
The bottom row of Figure~\ref{exampleA:fig1} shows the same
simulations
with the addition of a 
uniform random number between -0.1 and 0.1 added to
 each transmitted sample, representing transmission noise affecting
 the communication channel.
Comparing the solid traces in the upper and bottom rows it appears
that the channel noise negatively affects the transmission rate but,
as indicated in Section~\ref{sec:robustness}, closed-loop (practical) stability is preserved.

Figure~\ref{exampleA:fig1bis}
 proposes a comparison between the case 
$P_2 = 0.1$ (solid curves in all traces of Figures~\ref{exampleA:fig1} and~\ref{exampleA:fig1bis}) 
and the state-feedback policy in \cite{Tabuada07}.
To apply the results of \cite{Tabuada07} to this
example, we consider $V = x^T P_1 x$ with the same $P_1$ used by our
algorithm (defined above) and, from 
\eqref{eq:elazy_sys} we have that
$\dot{V} \leq -\alpha |x|^2 + \gamma |x||e|$
where $\alpha = 1$ and $ \gamma = 4.046$. Thus, following the notation in
\cite[Equation (8)]{Tabuada07}, we enforce a transmission of a sample
when $\gamma |e| \geq \sigma \alpha|x|$, for $\sigma = 0.9 < 1$
(which preserves
the decrease of $V$ along the solutions).
The plots clearly illustrate the reduced transmission rate of our approach as compared to
the policy based on \cite[Equation (8)]{Tabuada07}.
In particular, when the initial condition is balanced ($x(0,0) = [1\
1]^T$
in the top row of Figure~\ref{exampleA:fig1bis}), a reduced
transmission rate is already visible. However, the greatest advantage
is experienced with the unbalanced initial condition $x(0,0) = [10\
1]^T$ of the bottom row.
It should be also recalled that the reduced transmission rate that we
achieve is due to the fact that we restrict our attention to linear
control systems, while the work in \cite{Tabuada07} addresses
nonlinear systems for which more conservative bounds need to be used
in general.

Figure \ref{exampleA:fig2} represents the time evolution of
the synchronous policy from the estimated state $\hat{x}$ and
without any noise. The observer gain is 
$L = \smallmat{-14.77 \\ -6.68}$. For a fixed value of $P_2=0.1$, 
the convergence to zero depends on the initial 
mismatch between $\hat{x}$ and $x$. 
}

\subsection{Asynchronous policy with weight variations}

\begin{figure}[ht!]
\begin{center}
\subfigure[$P_2 = I$, $\alpha_1=0.9$, $\alpha_2=0.1$]{
	\includegraphics[width=0.9\columnwidth]{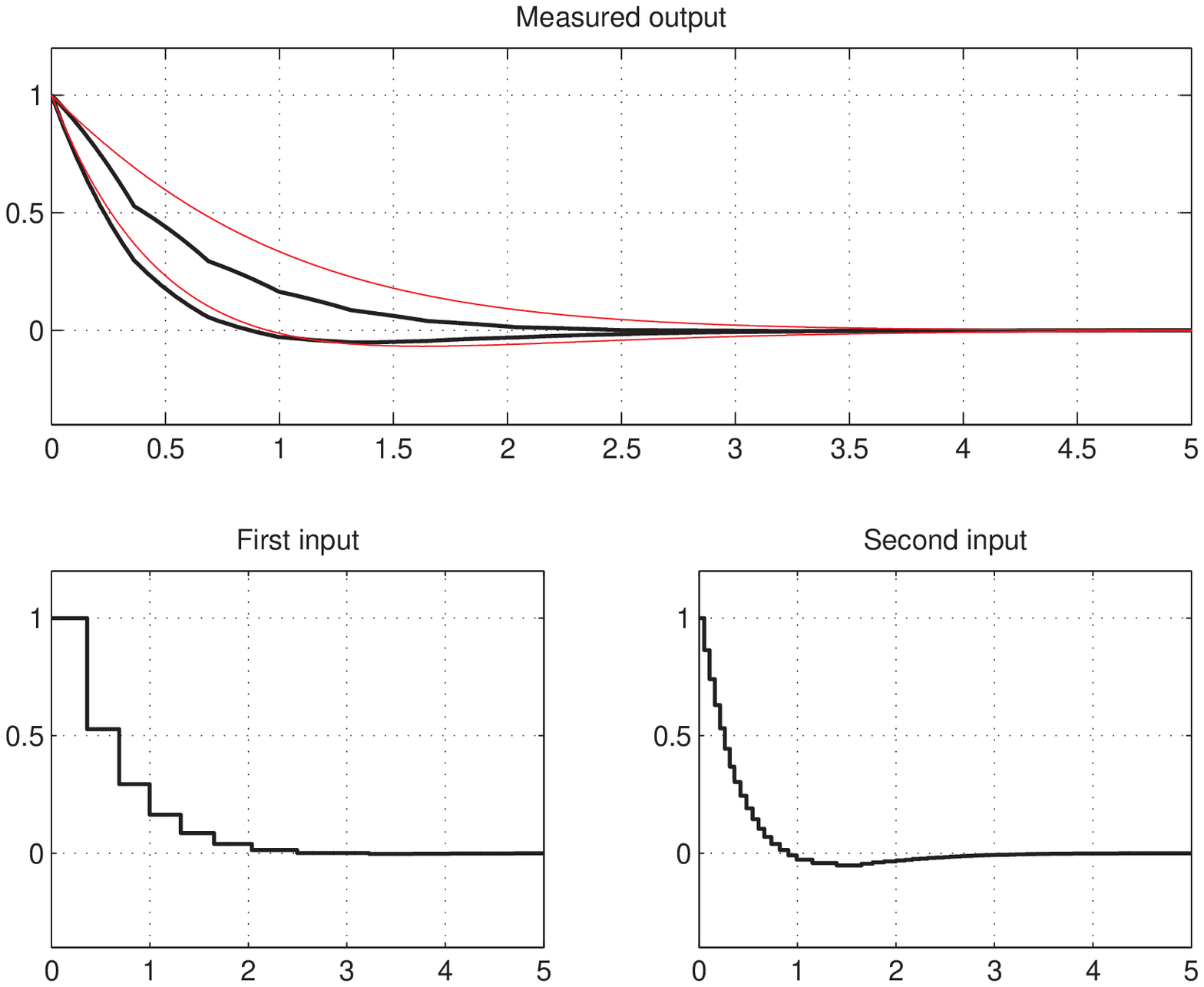}
  \label{fig:example2_P22_1_alpha_0.9} } 
\subfigure[$P_2 = I$, $\alpha_1=0.1$, $\alpha_2=0.9$]{
	\includegraphics[width=0.9\columnwidth]{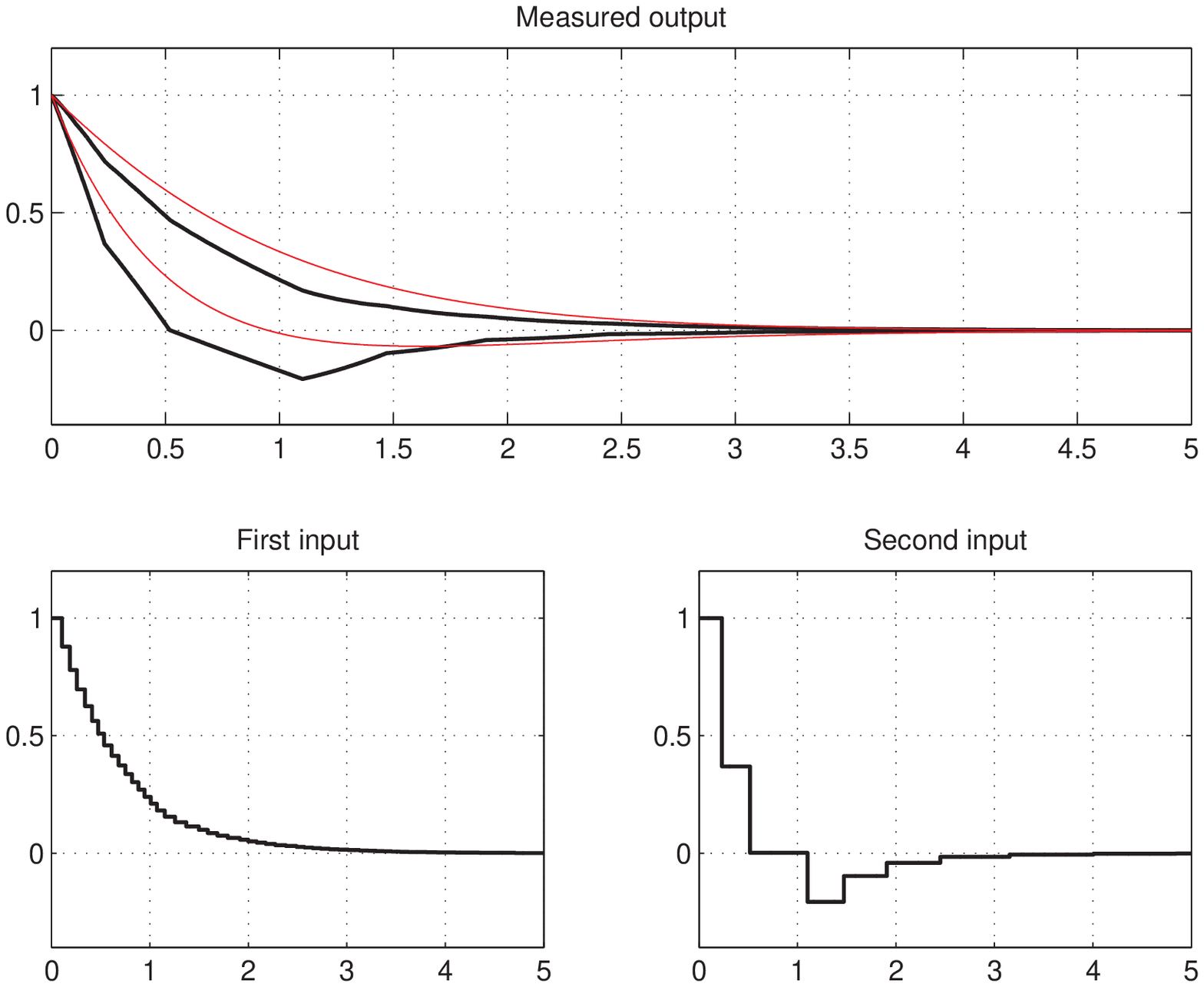}
  \label{fig:example2_P22_1_alpha_0.1} }
\caption{Asynchronous policy. Effects of variations on the vector $\alpha$.
The thin line in each figure is the output of the nominal closed-loop system.
} 
\label{figure_example2} 
\end{center}
\end{figure}

We consider the following unstable 
linear plant 
\begin{equation}
\label{plant_example1}
\left\{
\begin{array}{lll}
 \dot{x}_p & = & 
  \smallmat{ 1 & 1 \\ 0 & 1 } x_p + 
  \smallmat{ 1 & 0 \\ 0 & 1 } u_p \\
  y_p & = &\smallmat{ 1 & 0 \\ 0 & 1 }x_p.
\end{array}
\right. 
\end{equation}
which can be stabilized by the following LQR gains
\begin{equation}
\label{controller_example1}
 y_c =  \smallmat{ -2.1961 &  -0.7545 \\  -0.7545 &  -2.7146  } u_c
\end{equation}
through the interconnection $u_p = y_c$ and $u=y$.
The effect of the introduction of the t-lazy sensors operating
through the asynchronous policy is reported in Figure
\ref{figure_example2}.
Note that by choosing different $\alpha_1$ and $\alpha_2$,
we force one sensor to allow for a larger error 
bound on $e_i=\nu_i-y_i$ before transmitting, from which 
one sensor transmits its measurement $y_i$  more frequently than
the other one.

\section{Conclusions}
\label{sec:conclusions}
We introduced the transmission-lazy sensors to transform a continuous closed-loop system
to a system whose feedback signal is sampled and transmitted, possibly over a 
{ digital channel},
and we proposed two transmission policies which preserve the stability of 
the original closed-loop system. The first transmission policy requires an
update of the whole measured output vector $y$ based on a centralized decision,
while the second transmission policy allows for an asynchronous transmission,
in which each sensor decides its own transmission. Moreover, when the input matrix 
is full column rank, we showed that these policies guarantee global exponential stability.
Finally, by relying on an estimate of the state
from a classical continuous-time observer, 
both approaches have been extended to the case in which only the output 
of the plant-controller cascade is available.

\linespread{0.9}

\end{document}